\documentclass[format=acmsmall, review=false, screen=true]{acmart}

\usepackage{booktabs} 
\usepackage{multirow} 
\usepackage[english]{babel}
\usepackage[ruled]{algorithm2e} 

\SetAlFnt{\small}
\SetAlCapFnt{\small}
\SetAlCapNameFnt{\small}
\SetAlCapHSkip{0pt}
\IncMargin{-\parindent}

\usepackage{subfigure}
\usepackage{tikz}
\usetikzlibrary{trees}

\usepackage{color,colortbl}
\definecolor{blue}{RGB}{232,238,247}
\definecolor{green}{RGB}{204,255,204}
\definecolor{red}{RGB}{255,204,204}
\newcolumntype{b}{>{\columncolor{blue}}c}
\newcolumntype{g}{>{\columncolor{green}}c}
\newcolumntype{r}{>{\columncolor{red}}c}

\newcommand{\cat}[1]{{`#1'}}

\acmJournal{CSUR}
\acmVolume{52}
\acmNumber{6}
\acmArticle{131}
\acmYear{2019}
\acmMonth{12}
\setcopyright{acmcopyright}
\acmDOI{10.1145/3369782}
\received{October 2018}
\received[revised]{July 2019}
\received[accepted]{October 2019}
\begin{document}
\title[]{
Random Graph Modeling: A survey of the concepts
}

\author{Mikhail Drobyshevskiy}
\orcid{0000-0002-1639-9154}
\affiliation{%
  \institution{Ivannikov Institute for System Programming of the Russian Academy of Sciences and Moscow Institute of Physics and Technology}
  \streetaddress{109004, Moscow, Alexander Solzhenitsyn st., 25.}
  \city{Moscow}
  \country{Russia}}
\email{drobyshevsky@ispras.ru}

\author{Denis Turdakov}
\orcid{0000-0001-8745-0984}
\affiliation{%
  \institution{Ivannikov Institute for System Programming of the Russian Academy of Sciences and Moscow Institute of Physics and Technology}
  \streetaddress{109004, Moscow, Alexander Solzhenitsyn st., 25.}
  \city{Moscow}
  \country{Russia}}
\email{turdakov@ispras.ru}

\begin{abstract}

Random graph (RG) models play a central role in the complex networks analysis. They help to understand, control, and predict phenomena occurring, for instance, in social networks, biological networks, the Internet, etc.

Despite a large number of RG models presented in the literature, there are few concepts underlying them. Instead of trying to classify a wide variety of very dispersed models, we capture and describe concepts they exploit considering preferential attachment, copying principle, hyperbolic geometry, recursively defined structure, edge switching, Monte Carlo sampling, etc. We analyze RG models, extract their basic principles, and build a taxonomy of concepts they are based on. We also discuss how these concepts are combined in RG models and how they work in typical applications like benchmarks, null models, and data anonymization.

\end{abstract}

%
%
%
%

%
%



\maketitle



\section{Introduction}

\paragraph{Motivation to a network modeling}

Many real-world systems can be considered as networks of a set of connected discrete objects. Networks that demonstrate non-regular topological patterns are referred to as \textit{complex networks} which are the subject of intensive research in network science. They are used in a wide variety of areas of human activity: technological, social, biological, information.

Analysis of different aspects of complex networks is trying to answer important questions: How reliable the Internet is? What is the organization of social relations reflected in social networks? How diseases are spread, how information flows are distributed, and how to govern them?

These questions motivate creating realistic models of complex networks called \textit{random graphs}\footnote{Following~\cite{Kolaczyk:2009:SAN:1593430} we prefer to use term ``graph'' to refer to a mathematical abstraction of a real object while the term ``network'' corresponds to a more general sense of ``a collection of interconnected things''. In fact, these two words are usually used interchangeably in the literature.}, which help us to understand and control phenomena lying behind them. They attempt to find mechanisms of a network topology formation. For example, preferential attachment principle known as ``rich get richer'' was invented to explain scale-free property observed in many networks~\cite{barabasi1999emergence}.

The realism of the models is an important point of interest. For instance, we want to capture current patterns of the Internet to be able to study its evolution in the future. 

At the same time, a balance between realism and randomness should be provided. For instance, if one wants to preserve user privacy, simple relabeling of nodes in a social network does not protect from an adversary to learn whether an edge exists between two persons~\cite{backstrom2007wherefore}.

Random graphs are also important from a technical point of view. Many real networks exist in a few instances. However, we need scalable synthetic datasets for analysis. For instance, to test the significance of a new Facebook community detection algorithm, one needs a set of random graphs similar to the Facebook social graph. Another common scenario is to specify a null model and use it for hypotheses testing. For example, network motifs could be identified as subgraphs over-represented in the network compared to the null model~\cite{milo2002network}.

\paragraph{The survey focus}

The total number of RG models and generators is permanently growing. A single review is not able to cover all of them. Many modeling approaches exploit similar principles. Thus, they are very alike, while they may look different in some details. Literature reviews suffer from incompleteness by limiting themselves to particular applications. 

Instead of describing all RG models, we focus on the main concepts they use to achieve the goals. 
We noted that almost all approaches are based on a few numbers of high-level principles or concepts. Like building blocks, they are used in various combinations and modifications, giving a vast number of different algorithms for modeling random graphs. 
We systematically collect most known RG models, extract the basic principles they are based on, and classify them. Such a taxonomy gives a high-level RG overview and simplifies orientation in literature. 
Moreover, such concepts help researchers to design novel models and generators combining working elements in a new way.

Network modeling in the real world often goes beyond simple graphs. The nodes could have attributes, edges could be directed and weighted. 
Also, more specialiezed types of graphs are used, e.g., bipartite or multigraphs, hypergraphs (for communication in wireless networks)~\cite{avin2010radio}, and multilayer ones~\cite{boccaletti2014structure}. 
In this paper, we restrict ourselves on widely used directed weighted graphs with node labels.

\paragraph{Contributions}

Our contributions are threefold.
\begin{itemize}
\item First, we present a summary of recent efforts on random graph modeling guiding over monographs and notable reviews considering several topics of interest (Table~\ref{tab:review_of_reviews}).
\item Second, we present a taxonomy of concepts of graph modeling considering the hierarchical classification illustrated by particular models (figure~\ref{fig:taxonomy}).
\item Finally, we discuss applications of random graph models discussing the role of the described concepts.
\end{itemize}

The rest of the paper is organized as follows. In Section~\ref{sec:definition}, we clarify the terminology, recall the main graph metrics and features important for random graph modeling. In Section~\ref{sec:literature}, we provide the methodology of the literature analysis, our method for collecting relevant papers, guide of prominent reviews, and existing models classifications. Then, in Section~\ref{sec:taxonomy}, we present our taxonomy of the concepts, together with all constituents description with examples. In Section~\ref{sec:discussion}, we discuss the taxonomy and how the concepts work in various applications. Finally, we provide a conclusion and future work in Section~\ref{sec:conclusion}.

\begin{table}
\caption{Notations and acronyms}
\begin{tabular}{|c|l||c|l|}
\hline
$n$ & number of nodes & RG & random graph \\
\hline
$m$ & number of edges & DD & node degree distribution \\
\hline
$i$, $j$, $u$, $v$ & particular node & CC & clustering coefficient \\
\hline
$d_i$ & degree of node $i$ & ER & Erd\H{o}s-R\'{e}nyi model \\
\hline
$P_{ij}$ & edge $(i,j)$ probability & PA & Preferential attachment (principle) \\
\hline
$A_{ij}$ & adjacency matrix & MCMC & Markov chain Monte Carlo (sampling) \\
\hline
$\lambda_k$ & $k$-th eigenvalue of a matrix & & \\
\hline
\end{tabular}
\label{tab:notation}
\end{table}

\section{Concepts and Definitions}
\label{sec:definition}

We assume that the reader knows the main concepts of network science and is familiar with simple random graph models. Hence, we omit definitions of the basic terms and details of well-known models. Otherwise, we provide references to graphs and probability theory backgrounds~\cite{Kolaczyk:2009:SAN:1593430}. Notations and acronyms used throughout the paper are presented in Table~\ref{tab:notation}.

\subsection{Terminology Nuances}

Like in many fields, in graph related literature, the same things can be named by multiple terms. To clarify the terminology, we assume the following equalities, which generally hold in papers:

\begin{itemize}
\item "graph" = "network" = "network graph"
\item "node = "vertex", \;\; "edge" = "link" = "arc"
\item "random graph" = "random graph model" = "graph model" = "network model"
\item "graph feature" = "graph pattern"
\item "graph metric" = "graph measure"
\end{itemize}

On the other hand, we explicitly discriminate several concepts by using different terms to avoid misinterpretation.
                   
\paragraph{Graph model vs. graph generator}

We consider the graph model as a model in a mathematical sense. It specifies a description and conditions for a statistical object. A graph generator is an algorithm whose execution results in a (random) graph. Usually, the graph generator implements several RG model. Alternatively, a specification of the generator could be given first, which implicitly defines the RG model.

\paragraph{Graph metric/measure vs. graph feature/pattern}

The graph metric or measure refers to a function measuring the characteristics like diameter, degree distribution, adjacency matrix spectra. These terms are common in a quality estimation context.
Terms feature or pattern are used to speak about distinctive attributes of a particular graph. Both in qualitative manners (heavy tail degree distribution, high clustering) and quantitative ones (node degree sequence itself, diameter value).

\subsection{Random Graph Definitions}

What is called a random graph? In almost all cases it is meant by default the Erd\H{o}s-R\'{e}nyi (ER) model which refers to one of two very similar classic models: $G(n, m)$~\cite{erdds1959random} introduced by Erd\H{o}s and R\'{e}nyi, and $G(n, p)$ suggested by Edgar Gilbert~\cite{gilbert1959random}.
$G(n, m)$ gives equal probabilities to all graphs with $n$ nodes and $m$ edges, while in $G(n, p)$ each possible edge on $n$ nodes appears independently with constant probability $p$.
These two models are mostly used in applications and are extensively developed.

Actually, in literature, one can encounter diverse notions behind the term random graph. The following four citations exemplify this:
\begin{itemize}
\item "A network is said to be random when the probability that an edge exists between two nodes is completely independent of the nodes' attributes. In other words, the only relevant function is the degree distribution $P(k)$."~\cite{barrat2008}
\item "In full generality, by a random graph on a fixed number of vertices ($n$) we mean a random variable that takes its values in the set of all undirected graphs on $n$ vertices.
[...]
A random graph model is given by a sequence of graph valued random variables, one for each possible value of $n$: $\mathcal{M}=(G_n; n \in N)$."~\cite{farago2009structural}
\item "In general, a random graph is a model network in which a specific set of parameters take fixed values, but the network is random in other respects."~\cite{newman2010networks}
\item "[...] to specify a random N-node graph, we must give the set $G \subseteq \{0, 1\}^{N^2}$ of allowed graphs (the configuration space), together with a probability distribution $p(A)$ over this set. This combination $\{G, p\}$ of a graph set $G$ with associated probabilities is called a random graph ensemble.
Equivalently, we could always take $G = \{0, 1\}^{N^2}$ and assign $p(A) = 0$ to all disallowed graphs A."~\cite{coolen2017generating}
\end{itemize}

In general, "random graph" can refer to any model, wherein is specified a probability distribution over a set of graphs. For instance, E.\,D.\,Kolaczyk~\cite{Kolaczyk:2009:SAN:1593430} uses the notion of graph model as a collection $\{ \mathbb{P}_{\theta}(G), G \in \mathcal{G}, \theta \in \Theta \}$, where a parameterized probability space $\mathbb{P}_{\theta}$ is defined on an ensemble $\mathcal{G}$ of possible graphs.
There are two ways to express the complexity of a model: to incorporate it in $\mathbb{P}(G)$ specification or to restrict the set $\mathcal{G}$ of allowed graphs in a non-trivial way.
In the latter case, $\mathbb{P}(G)$ is typically assumed to be uniform, i.e., a generator would randomly pick a graph from $\mathcal{G}$, making the model more analytically tractable.
That is why the ER model is so popular and very well studied theoretically.

In this paper, we use the random graph equivalently to the graph model referring to a general case, where a mathematical construction defines a probability distribution over a set of possible graphs.

\subsection{Popular graph metrics and features}
\label{ssec:metrics}

During the history of network science, many graph patterns were discovered, and graph metrics were designed to measure their characteristics. Metrics help to discover new patterns in networks, which are analyzed to understand their nature. Most important features are in the focus of graph models, which try to explain their emergence.

To understand the properties of subnetworks, we quantitatively analyze the clustering properties, subgraph distribution, density distributions, and other metrics. We consider only the most notable graph metrics in the context of RG modeling. We start with static topological properties, then, ones describing graphs in dynamics and metrics related to the node and edge attributes. In this way, we underline the most popular patterns.

\begin{table}[h!]
\caption{Graph metrics and features. PL stands for power-law. $\alpha$ is used as PL exponent. Other notations are explained in Section~\ref{ssec:metrics}.}
\begin{center}
\begin{tabular}{|p{1.9cm}|p{2.8cm}|l|}
\hline
\textbf{Metric class} & \textbf{graph metric} & \textbf{frequently observable features} \\ \hline
\multirow{3}{*}{Degree} & DD & PL: $P(d) \sim d^{-\gamma}, \gamma \geqslant 1$, usually $\in (2;10)$. Sometimes DPLN \\ \cline{2-3}
 &  assortativity & $>0$ in social, $<0$ in biological and technological domains \\ \cline{2-3}
 & $dK$-distribution &  \\ \hline
\multirow{3}{*}{Subgraphs} & CC & much higher than in the ER model\\ \cline{2-3}
 & CC($d$) & PL \\ \cline{2-3}
 & subgraph distribution &  \\ \hline
\multirow{4}{*}{Connectivity} & effective diameter & small-world effect: $d_{eff}$ is small, often around 6 \\ \cline{2-3}
 & hop-plot & PL: $N_{pairs}(h) \sim h^{\alpha}$ for $h \ll d_{eff}$ \\ \cline{2-3}
 & connected components & presence of a giant connected component \\ \cline{2-3}
 & community structure & high modularity \\ \hline
\multirow{4}{*}{Spectral} & spectral radius & \\ \cline{2-3}
 & algebraic connectivity &  \\ \cline{2-3}
 & singular values of $A_{ij}$ & PL: $\lambda_k \sim k^{\alpha}$ for $k < n^{1/2}$, $\alpha \in (2;10)$ \\ \cline{2-3}
 & eigenvalues of Laplacian matrix & PL: $\lambda_k \sim k^{\alpha}$ for $k < n^{2/3}$, $\alpha \in (2;10)$ \\ \hline
\hline
\multirow{3}{*}{Dynamic} & $m(n)$ dependency & PL: densification $m(t) \propto n(t)^{\alpha}, \alpha \in (1;2)$ \\ \cline{2-3}
 & $d_{eff}(t)$ & shrinking over time \\ \cline{2-3}
 & properties over time & presence of gelling point \\ \hline
\hline
\multirow{3}{0pt}{Community labels} & community size & heavy-tailed distribution \\ \cline{2-3}
 & number of memberships of a node & heavy-tailed distribution \\ \cline{2-3}
 & $m_c(n_c)$ dependency & PL: densification $m_c \sim n_c^{\alpha}$ for communities $c$ in a network \\ \hline
\hline
\multirow{4}{*}{Edge weights} & total weight & PL: $W(t) \sim m(t)^w, w > 1$ \\ \cline{2-3}
 & node strength & PL: $s_i \sim d_i^w$ \\ \cline{2-3}
 & weighted principal eigenvalue & PL: $\lambda_1(t) \sim m(t)^{\beta}, \beta \in (0.5; 1.6)$ \\ \cline{2-3}
 & weight addition & self-similarity over time \\ \hline
\end{tabular}
\end{center}
\label{tab:metrics}
\end{table}

\subsubsection{Topology}

We group topology metrics into four classes reflecting their main aspects: node degrees, subgraphs, connectivity properties, and spectral features.

\paragraph{Node degrees}

Node degree is a basis for a set of important collective graph metrics: node degree distribution, node degree assortativity, and node degree correlations.

\begin{itemize}
\item Node degree distribution (DD). DD is one of the most famous characteristics, which counts nodes with a given number of neighbors. An important observation is that many real networks from various domains exhibit power-law DD~\cite{barabasi1999emergence}, for instance, $P(d) \sim d^{-\gamma}$ with various values of exponent $\gamma \geq 1$, commonly, between 2 and 10~\cite{eikmeier2017revisiting}. Independence of the scale parameters is called scale-free property. The same law holds for input (in-DD) and output (out-DD) links of directed graphs~\cite{ebel2002scale}. For some networks, like Mobile calls, a better fit could be a double Pareto-lognormal (DPLN) distribution~\cite{fang2012double}, a kind of a middle between Pareto and lognormal distributions.
\item Node degree assortativity. It is computed as the correlation between the node degree and average degree of its neighbors. The positive correlation is found in social networks: high degree nodes tend to connect to high degree nodes, while low degree nodes tend to connect to low degree nodes, which are referred to as assortative networks. Biological and technological networks are often disassortative with negative correlations~\cite{newman2002assortative}.
\item $dK$-distribution. $dK$-distribution shows the node degree correlation within subgraphs of size$d$ for arbitrary $d>0$~\cite{mahadevan2006systematic}. For $d=0$, it shows the average node degree $\langle d_i \rangle$. $d=1$ corresponds to the classical DD and $d=2$ corresponds to joint degree distribution $P(d_i, d_j)$. $d>2$ combines joint distributions for each possible (connected) edge configuration on $d$ nodes. Series of $dK$-distribution with increasing $d$ describe more complex features of a given graph becoming the complete one when $d=n$.
\end{itemize}

\paragraph{Subgraphs}

It is very useful to count triads (a combination of three nodes) and higher order substructures in graphs. Three characteristics are considered: clustering coefficient, clustering coefficient as a function of node degree, and subgraphs distribution.

\begin{itemize}
\item Clustering coefficient (CC). CC is the ratio of the number of closed triads (triangles) to the number of all triads. The transitivity coefficient is the clustering coefficient measured for the whole graph. The average local clustering coefficient is measured for each node and averaged over all nodes. It is found that in real networks, CC is significantly higher than if node pairs are linked independently like in ER model.
\item Clustering coefficient as a function of node degree. For some networks, clustering coefficient follows a power-law, which is associated with a hierarchical structure~\cite{costa2007characterization}.
\item Subgraphs distribution. Distribution of small subgraphs of size 3 or 4 could serve in two ways. As a feature vector, it contains enough information to categorize graphs over domains with high precision~\cite{bordino2008mining}. Detecting statistically significant subgraphs for a particular graph, called network motifs, could reveal network building principles. It is especially fruitful in a biological domain~\cite{milo2002network}.
\end{itemize}

\paragraph{Connectivity}

Distances in graphs give a picture of their global connectivity (like the effective diameter), reachability of nodes, connected components, and community structure.

\begin{itemize}
\item Effective diameter. While the diameter of the graph is the maximal distance between its nodes, the effective diameter $d_{eff}$ is a major fraction (typically 90 \%) of node pairs connected with at most $d_{eff}$ edges. It has a more informative feature than the diameter. For instance, it shows that social graphs and WWW have a low effective diameter (around 6), which is coined as 'small-world' effect~\cite{watts1998collective}.
\item Hop-plot. For a given path length $h$, it shows how many node pairs are reachable in $h$ hops. Hop-plot demonstrates the shortest path length distribution in the graph. This metric aggregates two related characteristics, including average shortest path length and effective diameter. Faloutsos et al.~\cite{faloutsos1999power} observed that the Internet demonstrates hop-plot exponent: number of node pairs is proportional to a power of $h$ for $h \ll d_{eff}$.
\item Connected components. Typically, a network is a connected graph that contains one large connected component. Thus, an important question concerns the appearance of a giant connected component in a random graph (phase transition)~\cite{erdos1960evolution}, which is related to percolation theory.
\item Community structure. The presence of tightly connected groups of nodes is observed in social networks, where they reflect groups of interest. In biological networks, they correspond to the functional groups. The knowledge of how well community structure is expressed in a graph is given by modularity measure~\cite{newman2006modularity}. The communities are characterized by additional topological metrics like conductance, separability, and cohesiveness~\cite{yang2015defining}.
\end{itemize}

\paragraph{Spectra}

Graph features are tightly connected to its spectral properties: eigenvalues, eigenvectors of its adjacency and Laplacian matrices. The spectral analysis is used to study processes on networks and develop algorithms on graphs.
For example, Google search engine is based on the Perron-Frobenius eigenvector of the web graph. In general, this is the subject of graph spectral theory~\cite{brouwer2011spectra}. 
We consider spectral radius, algebraic connectivity, singular value distribution of the adjacency matrix, and eigenvalue distribution of the Laplacian matrix as a keys of spectra classification.

\begin{itemize}
\item Spectral radius. The maximal eigenvalue $|\lambda_1|$ of the graph adjacency matrix is called its spectral radius. $|\lambda_1|=0$ corresponds to a disconnected graph. Thus, the spectral radius is usually computed for its giant component. Spectral radius does not increase when nodes or edges are removed from the graph. It serves as an alternative size metric. For instance, it is shown that the smaller radius means the higher robustness to virus spreading~\cite{van2009virus}.
\item Algebraic connectivity. The second smallest nonzero eigenvalue of graph Laplacian matrix $L$ is called algebraic connectivity. It is also measured for a giant component. It is the larger the better graph is connected. An eigenvector, corresponding to this eigenvalue, is called Fiedler vector. It is useful for graph partitioning~\cite{pothen1990partitioning}.
\item Singular value distribution of the adjacency matrix. It was found that it follows the power law in real networks~\cite{eikmeier2017revisiting}. This law often holds for $n^{1/2} - n^{2/3}$ largest singular values.
\item Eigenvalue distribution of the Laplacian matrix. Top $k$ eigenvalues follow power law distribution $\lambda_k \sim k^{\alpha}$, where $k$ scales as $n^{2/3}$ and $\alpha$ usually varies in $(2; 10)$~\cite{eikmeier2017revisiting}. It was noted, that the exponent of this power law is often nearly identical to the DD power law exponent, for graphs where these power laws were statistically significant.
\end{itemize}

There are other graph metrics, helpful in their evaluation but not playing a significant role in designing RG models, such as resilience and principal eigenvector. For a more complete survey of graph metrics, we refer to Costa, L. D. F. et al.~\cite{costa2007characterization}, and Hern\'{a}ndez, J. M., and Van Mieghem, P.~\cite{hernandez2011classification}.

\subsubsection{Dynamics}

Many real graphs evolve in time, showing appearance and disappearance of new nodes and edges. In practice, most networks grow, i.e., the number of nodes $n$ increases over time $t$. All known static topology metrics can be measured through time variable, as well as their mutual dependence, which reveals new dynamical graph patterns. Further, we consider densification power law, shrinking diameter, and gelling point.

\begin{itemize}
\item Densification power law. The number of edges $m$ grows as a power of the number of nodes: $m(t) \sim n(t)^{\alpha}$, where $1 < \alpha < 2$~\cite{leskovec2007graph}.
\item Shrinking diameter. In many cases, the effective diameter is decreased with network growth~\cite{leskovec2007graph}.
\item Gelling point. Real graphs have a moment of stabilization ('gelling') during their evolution, where diameter has a spike. After that moment diameter starts to shrink and other laws are obeyed: densification power law is satisfied well, the second and the third connected component begin oscillating around some constant values~\cite{mcglohon2008weighted}.
\end{itemize}

\subsubsection{Attributes}

Real networks contain a lot of information besides the topology. Nodes often have attributes: user profile data in a social network, protein properties, etc. Edges can also be labeled with timestamps, weights, and so on. We consider only node communities and edge weights.


\paragraph{Community labels}

Social networks are known to have an explicit community structure formed by users' attributes.
Such 'ground-truth' communities have common features despite they are from very different domains~\cite{yang2014structure}, for instance:
\begin{itemize}
\item heavy-tailed distribution of community size;
\item heavy-tailed distribution of the number of community memberships of a node;
\item densification power law: in the scope of one network, the number of edges in a community $c$ grows as a power of its size, $m_c \sim n_c^{\alpha}$.
\end{itemize}

Other properties are also important for community structure modeling. The probability of edge $P_{ij}$ is increased with increasing a number of common communities for $i$ and $j$. Nodes in the community overlaps are more densely connected than nodes in non-overlapping parts of communities; and others.


\paragraph{Edge weights}

Edge weights usually express the strength of connections between nodes. For example, they correspond to the number of word co-occurrences in a text, amount of network traffic, and indicate the presence of multiple edges, e.g., number of citations. Their properties can be described by a Weight power law, Snapshot power law, Weighted principal eigenvalue power law, Self-similarity, etc.

\begin{itemize}
\item Weight power law. A total edges weight grows as a power of the number of edges: $W(t) \sim m(t)^w$ with exponent $w > 1$~\cite{mcglohon2008weighted}.
\item Snapshot power law. Node strength $s_i$, defined as a total weight of its adjacent edges, depends on its degree $d_i$ as a power law: $s_i \sim d_i^w$. This holds when measured for incoming and outgoing edges separately~\cite{mcglohon2008weighted}.
\item Weighted principal eigenvalue power law. Largest eigenvalue of the weighted adjacency matrix grows as a power of the number of edges: $\lambda_1(t) \sim m(t)^{\beta}$, where exponent $\beta$ was observed to be $0.5 - 1.6$~\cite{mcglohon2008weighted}.
\item Self-similar weight addition. The rate of weight addition over time shows self-similarity~\cite{mcglohon2008weighted}.
\end{itemize}

A summary of the described metrics and features is presented in table~\ref{tab:metrics}. It shows that ten power laws are observed in real networks.

\section{Literature analysis}
\label{sec:literature}


In this section, we describe a method for retrieving relevant papers. Then, we analyze the most prominent review works and give several classification schemes of RG models.

When performing a literature search, we discover a dozen of large volume studies of RG models, which we describe in this section.
Firstly, we present our method for papers collecting, summarize various aspects of the most prominent reviews, and, finally, discuss the classifications of RG models.

\subsection{Papers collecting procedure}

Among a huge number of publications, we distinguish three types of papers of interest, with decreasing priority:

\begin{enumerate}
\item \textbf{reviews}: reviews and comparative studies of RG models;
\item \textbf{novelties}: works suggesting a new approach or extending an existing one;
\item \textbf{applications}: works applying an existing RG model to a particular problem.
\end{enumerate}

During the search process, we found that the last class is too vast to be analyzed manually. While there are tens of reviews and hundreds of new RG models, the amount of applications is much larger. Therefore, we concentrate on review papers and detecting most prominent works from the second class.

\paragraph{Databases querying}

We consider three databases as publication sources: Google Scholar, ACM Digital Library, and IEEE Xplore Digital Library. We start with a collection of already known to us papers (\cite{
leskovec2008statistical,
costa2007characterization,
toivonen2009comparative,
sala2010measurement,
gilbert2009social,
bonato2009survey,
lazzarin2011,
pofsneck2012physical,
leskovec2007graph,
chakrabarti2004r,
leskovec2005realistic,
leskovec2010kronecker,
palla2010multifractal,
nickel2008random,
akoglu2009rtg,
krioukov2010hyperbolic,
zhang2016gscaler,
seshadri2008mobile,
wegner2011random,
lancichinetti2009benchmarks,
chykhradze2014distributed,
mahadevan2006systematic,
ying2009graph,
staudt2016generating,
drobyshevskiy2017learning,
park2017trilliong}
)
and iteratively extend it with results obtained by querying mentioned databases.

For Google Scholar, we merge the results from the follows queries (option "Sort by relevance" is enabled):
\begin{itemize}
\item query \texttt{"(random OR artificial OR synthetic OR model OR modeling) (graph OR graphs OR network OR networks)"}. We select the first 150 papers;
\item query \texttt{"(random OR artificial OR synthetic OR model OR modeling OR modelling) (graph OR graphs OR network OR networks) (generation OR generating OR generator))"}. We select the first 130 papers;
\item query \texttt{"(random OR artificial OR synthetic OR model OR modeling OR modelling) (graph OR graphs OR network OR networks) (generation OR generating OR generator OR generative))"}. We select "since 2009", "since 2013" and "since 2016" and take 50 relevant papers from each result.
\end{itemize}

Unfortunately, despite queries variability, the search results may still miss eligible works, but include many irrelevant ones. The number of first papers was chosen as a trade-off.

For ACM Digital Library, we run queries:
\begin{itemize}
\item "any field" matches all: \textit{random graph network model generation}. We sort by relevance and select the first 50 papers;
\item "abstract" matches all: \textit{random graph network model generation}. We sort by relevance and select the first 50 papers;
\item "abstract" matches all: \textit{"random graphs" network model generator}, and "abstract" matches any: \textit{review survey overview comparison}. We sort by relevance and select the first 30 papers;
\end{itemize}

For IEEE Xplore Digital Library, we perform searches in metadata, and select 10, 32, and 33 papers from three corresponding results:
\begin{itemize}
\item \textit{"random graph" AND network AND model AND generator};
\item \textit{"random graph" AND network AND model AND generation};
\item \textit{"random graph" AND network AND model AND generating}.
\end{itemize}

Google Scholar indexes most publications of the interest and returns the most relevant papers. We extracted around 300 papers. ACM and IEEE databases additionally contributed 70 and 46 papers, respectively.

To complete the review papers class, we retrieve reviews and scan links they contain to find other reviews. We eliminate works written earlier than 15 years ago (before 2003), except most valuable publications like Erd\H{o}s-R\'{e}nyi's and Mark Newman's ones.

Also, we added the results of similar queries to Google Books. 
We completed our collection with occasional relevant papers encountered during our analysis.

\subsection{Review of reviews}

In the last 15 years, the most extensive study was presented in monographs~\cite{
dorogovtsev2003evolution,
penrose2003random,
newman2011structure,
durrett2007random,
caldarelli2007,
alessandro2007large,
bonato2008course,
barrat2008,
Kolaczyk:2009:SAN:1593430,
newman2010networks,
lovasz2012large,
raigor2012models,
Chakrabarti2012,
harris2013introduction,
van2014random,
frieze2015introduction,
coolen2017generating}.
Reviews and comparisons of random graph models were conducted in works~\cite{
newman2003structure, chakrabarti2006graph, 
toivonen2009comparative, farago2009structural,
gilbert2009social, bonato2009survey,
goldenberg2010survey, sala2010measurement, 
lazzarin2011, pofsneck2012physical,
raigor2012models, amblard2015models, meyer2017large}.

To make an overview of large volume issues, we analyzed how they reveal our topics of interest. Table~\ref{tab:review_of_reviews} is a quick guide of what information one can find in which books (covers only \textit{large volume} issues). Topics of interest and why they are important for RG modeling is described further.

\paragraph{RG models/generators description}

RG models and generators are our main focus. Each of the considered publications describes models.
Much attention to various models is paid in \cite{durrett2007random, alessandro2007large, newman2010networks, Chakrabarti2012, bernovskiy2012random, frieze2015introduction, coolen2017generating}. Book of M.\,Penrose~\cite{penrose2003random} is fully devoted to random geometric graphs, J. K. Harris~\cite{harris2013introduction} focuses on exponential random graph models.

\paragraph{RG models/generators classification}

The number of models and generators suggested is counted in hundreds or even thousands, so we want a more general view of them. Several classifications of existing approaches could be more informative than the details of a particular model.

We did not find an exhaustive taxonomy of RG models in the literature. Most of them are out-of-date or suffer from incompleteness. Since there is no conventional classification of RG models, each work suggests its state-of-art view. The most detailed classifications are provided in works~\cite{toivonen2009comparative, goldenberg2010survey, Chakrabarti2012, bernovskiy2012random, amblard2015models}, while D.\,Chakrabarti and C.\,Faloutsos~\cite{Chakrabarti2012} give a table with 24 RG generators compared by several graph metrics.
Alternative arrangements can be found in~\cite{farago2009structural, alessandro2007large, barrat2008, Kolaczyk:2009:SAN:1593430, newman2010networks, Chakrabarti2012, coolen2017generating}.

\paragraph{Networks examples / classification}

Network data come from many sources and can be differentiated by research domains (e.g., society, biology) and graph specificity, i.e., large, small, directed, weighted, with metadata, etc.
Each network domain rises its specific problems, which makes individual requirements to RG models exploited in it. For example, a biological graph with a hundred of nodes and a social graph with millions of nodes and billions of edges require different modeling approaches and impose specific constraints.

Traditionally, authors distinguish from 4 up to 10 network domains, see~\cite{newman2003structure, dorogovtsev2003evolution, caldarelli2007, barrat2008, newman2010networks}. Subdomains could also be introduced, e.g., Konect database of networks by J.\,Kunegis~\cite{kunegis2013konect} contains 24 categories, but no fixed hierarchy is generally accepted.
%
%

\paragraph{Network metrics, patterns}


A big number of real-world networks from different domains appeared to have common patterns with similar characteristics: power law of degree distribution, small diameter, high clustering, etc.~\cite{albert2002statistical}. These features are extremely represented in practice, but not intrinsic to classical ER graphs. Therefore, we need RG models with such properties.

In the context of RG modeling, the knowledge of network patterns can serve in several ways:

\begin{itemize}
\item reproducing specific network patterns makes RG models more realistic;
\item graph metrics allow to compare corresponding network patterns and evaluate RG models' quality;
\item better understanding of a network object, e.g., network motifs reflects behavior patterns in biological networks.
\end{itemize}

Large observations of network properties are given in studies~\cite{Kolaczyk:2009:SAN:1593430, newman2010networks, Chakrabarti2012} and \cite{dorogovtsev2003evolution, caldarelli2007, barrat2008, newman2003structure}; Bonato's book~\cite{bonato2008course} is fully devoted to the Web graph.

\paragraph{RG applications description}

Applications of networks are the main goal of RG modeling activity. They dictate requirements, conditions, and restrictions on RG models. RG applications can be viewed in a dual way. First, each graph domain has specific typical tasks. For example, T.\,Coolen et al.~\cite{coolen2017generating} consider tasks arising in 5 domains: power grids, social networks, food webs, world wide web, and protein-protein interactions.

Second, a certain type of problems can appear in multiple domains. This point of view is more suited for works~\cite{newman2011structure, barrat2008}.

\begin{table}[h!]
\caption{Review of large volume issues over last 15 years. Legend: '-' - not covered, '1' - the topic is concerned slightly, '2' - topic is covered, '3' - a detailed survey, 's' - special focus.}
\begin{center}
\begin{tabular}{|p{2.4cm}|c|c|c|c|c|c|c|c|c|c|c|c|c|c|c|c|c|c|}
\hline
\textbf{Topic covered} & 
\rotatebox[origin=l]{90}{Dorogovtsev \& Mendes \cite{dorogovtsev2003evolution}} &
\rotatebox[origin=l]{90}{Penrose \cite{penrose2003random}} &
\rotatebox[origin=l]{90}{Newman, Barabasi, Watts \cite{newman2011structure}} &
\rotatebox[origin=l]{90}{Durrett \cite{durrett2007random}} &
\rotatebox[origin=l]{90}{Caldarelli \cite{caldarelli2007}} &
\rotatebox[origin=l]{90}{Vespignani, Caldarelli \cite{alessandro2007large}} &
\rotatebox[origin=l]{90}{Bonato \cite{bonato2008course}} &
\rotatebox[origin=l]{90}{Barrat \cite{barrat2008}} &
\rotatebox[origin=l]{90}{Kolaczyk \cite{Kolaczyk:2009:SAN:1593430}} &
\rotatebox[origin=l]{90}{Newman \cite{newman2010networks}} &
\rotatebox[origin=l]{90}{Raigorodsky \cite{raigor2012models}} &
\rotatebox[origin=l]{90}{Lov{\'a}sz \cite{lovasz2012large}} &
\rotatebox[origin=l]{90}{Chakrabarti \cite{Chakrabarti2012}} &
\rotatebox[origin=l]{90}{Van Der Hofstad \cite{van2014random}} &
\rotatebox[origin=l]{90}{Frieze, Karo\`{n}ski \cite{frieze2015introduction}} &
\rotatebox[origin=l]{90}{Coolen, Annibale, Roberts \cite{coolen2017generating}} \\
\hline
year & \rotatebox[origin=l]{90}{2003} & \rotatebox[origin=l]{90}{2003} & \rotatebox[origin=l]{90}{2006} & \rotatebox[origin=l]{90}{2007} & \rotatebox[origin=l]{90}{2007} & \rotatebox[origin=l]{90}{2007} & \rotatebox[origin=l]{90}{2008} & \rotatebox[origin=l]{90}{2008} & \rotatebox[origin=l]{90}{2009} & \rotatebox[origin=l]{90}{2010} & \rotatebox[origin=l]{90}{2011} & \rotatebox[origin=l]{90}{2012} & \rotatebox[origin=l]{90}{2012} & 
\rotatebox[origin=l]{90}{2014} & \rotatebox[origin=l]{90}{2015} & \rotatebox[origin=l]{90}{2017} \\ \hline
RG models description & 2 & s & 2 & 3 & 2 & 3 & 2 & 2 & 2 & 3 & 3 & 1 & 3 & 2 & 3 & 3 \\ \hline
RG models classification & 1 & - & - & 1 & - & 1 & - & 2 & 2 & 2 & 1 & - & 2 & - & - & 2 \\ \hline
networks examples / classification & 3 & - & 1 & - & 3 & 2 & s & 2 & 2 & 3 & - & 1 & - & 1 & - & - \\ \hline
networks metrics, patterns  & 2 & - & 1 & - & 2 & 2 & s & 2 & 3 & 3 & - & - & 3 & 1 & - & 1 \\ \hline
RG applications described & - & - & 2 & - & 1 & - & - & 2 & 1 & - & 1 & - & - & - & - & 3 \\ \hline
algorithms and processes on networks & - & - & 1 & - & - & 3 & 2 & 3 & 2 & 3 & - & - & 1 & 1 & 1 & - \\ \hline
exercises & - & 1 & - & - & - & - & 3 & - & 2 & 2 & - & 1 & - & 2 & 3 & 2 \\ \hline
theoretical preliminaries & 3 & 1 & - & - & 2 & 2 & 2 & 1 & 3 & 2 & 3 & 1 & 1 & 2 & 1 & 2 \\ \hline
mathematical results & 2 & 3 & 2 & 3 & 2 & 1 & 3 & 2 & 1 & 2 & 3 & s & - & 3 & 3 & 3 \\ \hline
datasets described & - & - & - & - & - & - & - & - & 1 & - & - & - & 1 & - & - & - \\ \hline

\end{tabular}
\end{center}
\label{tab:review_of_reviews}
\end{table}


We can further follow several other topics, often covered in the network literature, but less relevant to RG modeling. They are represented in Table~\ref{tab:review_of_reviews}.

\paragraph{Algorithms and processes on networks}

RG models are used to develop and test various algorithms and processes on networks. There is no strict border between processes and algorithms. We try to separate them by examples. Examples of algorithms:

\begin{itemize}
\item network topology inference: link prediction; inference of association networks; tomographic network topology inference;
\item graph mining: community detection, modularity calculation; page rank, etc.
\end{itemize}

Process on a network is characterized with random variables $X_i$ (static) or $X_i(t)$ (dynamic), defined at nodes.
Examples of processes:

\begin{itemize}
\item static: nearest neighbor prediction; Markov random fields; kernel-based regression;
\item dynamic: virus spread, epidemic modeling, information; network flow (traffic), etc.
\end{itemize}

A lot of research on algorithms on networks and processes is contained in works~\cite{alessandro2007large, barrat2008, newman2010networks, Kolaczyk:2009:SAN:1593430}.

\paragraph{Mathematical results}

One of the research directions is the theoretical study of RG models and generators. Several graph models are well-studied due to their popularity and mathematical tractability, e.g., percolation theory for the ER model. Properties of Kronecker graph generators are extensively explored, and many extensions and modifications to the original model are developed.

Richest mathematical results are presented in works~\cite{penrose2003random, durrett2007random, lovasz2012large, van2014random, frieze2015introduction, coolen2017generating}.

\paragraph{Theoretical preliminaries}

For a non specialist in the field, it is important, whether the work gives a detailed introduction to the field. It is a kind of "barrier to entry" for the paper.

An introduction in graph theory is presented in~\cite{dorogovtsev2003evolution, caldarelli2007, alessandro2007large, newman2010networks, coolen2017generating}, while works~\cite{bonato2008course,
Kolaczyk:2009:SAN:1593430, raigor2012models, van2014random} provide also mathematical preliminaries.

\subsection{Existing classifications of random graph models}

A few works on RG modeling give an explicit classification of existing models. Moreover, usually, they consider only several categories of models popular in a chosen field of interest, e.g., social networks or biology, thus suffer from incompleteness.

Usually, RG modeling approaches consider two classes: static and dynamic. In static models the number of nodes $n$ is fixed and then are defined according to some rules based on nodes' attributes if specified. A straightforward example is the ER model. Dynamic models assume that nodes and edges are added iteratively depending on the current state of the graph, e.g., preferential attachment process. 
A separate class is constituted by Exponential Random Graph Models (ERGM), where they are defined by sets of conditions of graph statistics.

In this section, we consider several classifications covering RG models, and discuss social graph models since they belong to the most wide-spread domain.

\subsubsection{General models}

Leaving out domain-specific models, the majority of popular classification schemes~\cite{barrat2008, Kolaczyk:2009:SAN:1593430, goldenberg2010survey, newman2010networks} can be roughly reduced to the following:
\begin{enumerate}
\item Static models (also called equilibrium):
  \begin{itemize}
  \item ER models usually referred to as "random";
  \item generalized DD models.
  \end{itemize}
\item Dynamic models (also called growth, evolving, and non-equilibrium):
  \begin{itemize}
  	\item PA and its extensions;
  	\item copy and duplication models;
  	\item optimization-based models.
  \end{itemize}
\item Other models:
  \begin{itemize}
	\item ERGM;
  	\item small-world models.
  \end{itemize}
\end{enumerate}

However, it is useful to look at other classifications that do not fit into this scheme.
A good approach to a taxonomy of graph generators is given by D.\,Chakrabarti and C.\,Faloutsos~\cite{Chakrabarti2012}. The authors suggest five categories: 
\begin{enumerate}
\item Random Graph generators --- connect nodes using random probabilities;
\item Preferential Attachment generators --- give preference to nodes with more edges;
\item Optimization-based generators --- minimizing risks under limited resources leads to power law;
\item Geographical models --- nodes' geography affects network growth and topology;
\item Internet-specific generators --- hybrids of concepts to fit special features of the Internet.
\end{enumerate}

An alternative view on RG is developed by T.\,Coolen, A.\,Annibale, and E.\,Roberts~\cite{coolen2017generating}.
They consider graph ensembles, which are imposed by hard and soft constraints:\begin{enumerate}
\item graphs with constraints:
  \begin{enumerate}
  \item soft constraints --- graphs must have the chosen features on average (same as ERGM);
  \item hard constraints --- each graph must have the chosen features.
  \end{enumerate}
\item graphs defined by algorithms:
  \begin{enumerate}
  \item network growth algorithms (PA and extensions);
  \item specific models: small-world, geometric, planar, and weighted.
  \end{enumerate}
\end{enumerate}

\subsubsection{Social network models}
Social network models are very demanded and widely developed branch of complex network modeling. R.\,Toivonen et al.~\cite{toivonen2009comparative} suggest the following taxonomy which fits well in a generalized scheme:
\begin{enumerate}
\item network evolution models --- links addition depends on local network structure
  \begin{itemize}
	\item growing --- nodes are added until a certain size is reached;
	\item dynamical --- number of nodes is fixed, evolution continues until certain statistics stop to change.
  \end{itemize}
\item nodal attribute models --- link probability depends only on nodes' attributes (homophily, like to like, spatial model).
\item ERGM.
\end{enumerate}

F.\,Amblard and co-authors~\cite{amblard2015models} examine social network models presented in Journal of Artificial Societies and Social Simulation over 17 years (up to 2015) and sorted them into 9 categories:
\begin{enumerate}
\item Regular lattices;
\item Random networks --- mainly ER;
\item Small-world networks;
\item Scale-free networks --- mainly PA;
\item Spatial networks --- built from the spatial distribution of the agents using a distance;
\item Hierarchical structures --- tree-like graphs for organizational structures or familial networks;
\item Kinship networks --- bipartite graphs for the familial network;
\item Empirical networks --- empirical data on social networks are used to generate ones;
\item Other kind of models --- ad hoc models that strictly follow the modeled system.
\end{enumerate}

M.\,Bernovskiy  and N.\,Kuzyurin~\cite{bernovskiy2012random} suggest classification based on model complexity, although consider a limited number of models:
\begin{enumerate}
\item random graphs --- ER and its extensions;
\item simplest scale-free models --- Bollob{\'a}s model~\cite{bollobas2003directed} and extensions; copying model, etc.;
\item more flexible scale-free models --- generalized DD models (Chung-Lu~\cite{aiello2000random}, Janson-{\L}uczak~\cite{janson2010large}, etc).
\end{enumerate}

and on a partition of scale-free models:
\begin{enumerate}
\item fixed exponent --- power law DD and other properties are mathematically proved: Bollob{\'a}s-Riordan~\cite{bollobas2003directed} and extensions;
\item tunable exponent --- power law exponent is tunable, which allows for phase transitions research: Chung-Lu~\cite{aiello2000random}, Janson-{\L}uczak~\cite{janson2010large};
\item unknown properties --- properties are not yet proved: Forest Fire~\cite{leskovec2007graph} and others.
\end{enumerate}

An interesting focus is presented by A.\,Sala et al.~\cite{sala2010measurement}, where the authors split 6 models into 3 categories based on the methodology:
\begin{enumerate}
\item Feature-driven models --- reproducing statistical features of a graph: Barab{\'a}si-Albert~\cite{albert2002statistical}, ForestFire~\cite{leskovec2007graph};
\item Intent-driven models --- emulating the creation process of the original graph: Random Walk~\cite{vazquez2003growing}, Nearest Neighbor~\cite{vazquez2003growing};
\item Structure-driven models --- capturing statistics from the graph structure to reproduce it: Kronecker Graphs, $dK$-graphs.
\end{enumerate}

In the screened literature, we did not find a satisfactory overview of the existing RG models. All the attempts were out of date or far from completeness. As we see, there exist several classifications from different perspectives: whether the graph is growing or not, algorithm complexity, used methodology, from the application point of view, etc. However, low-level concepts working in models are still not clear. 
In the current paper, we review such simple basic mechanisms detected in the models. Further, we describe our vision of the area and give a comprehensive taxonomy of concepts.

\section{A taxonomy of random graph modeling approaches}
\label{sec:taxonomy}

\begin{figure}[h!]
\centering
\includegraphics[scale=0.52]{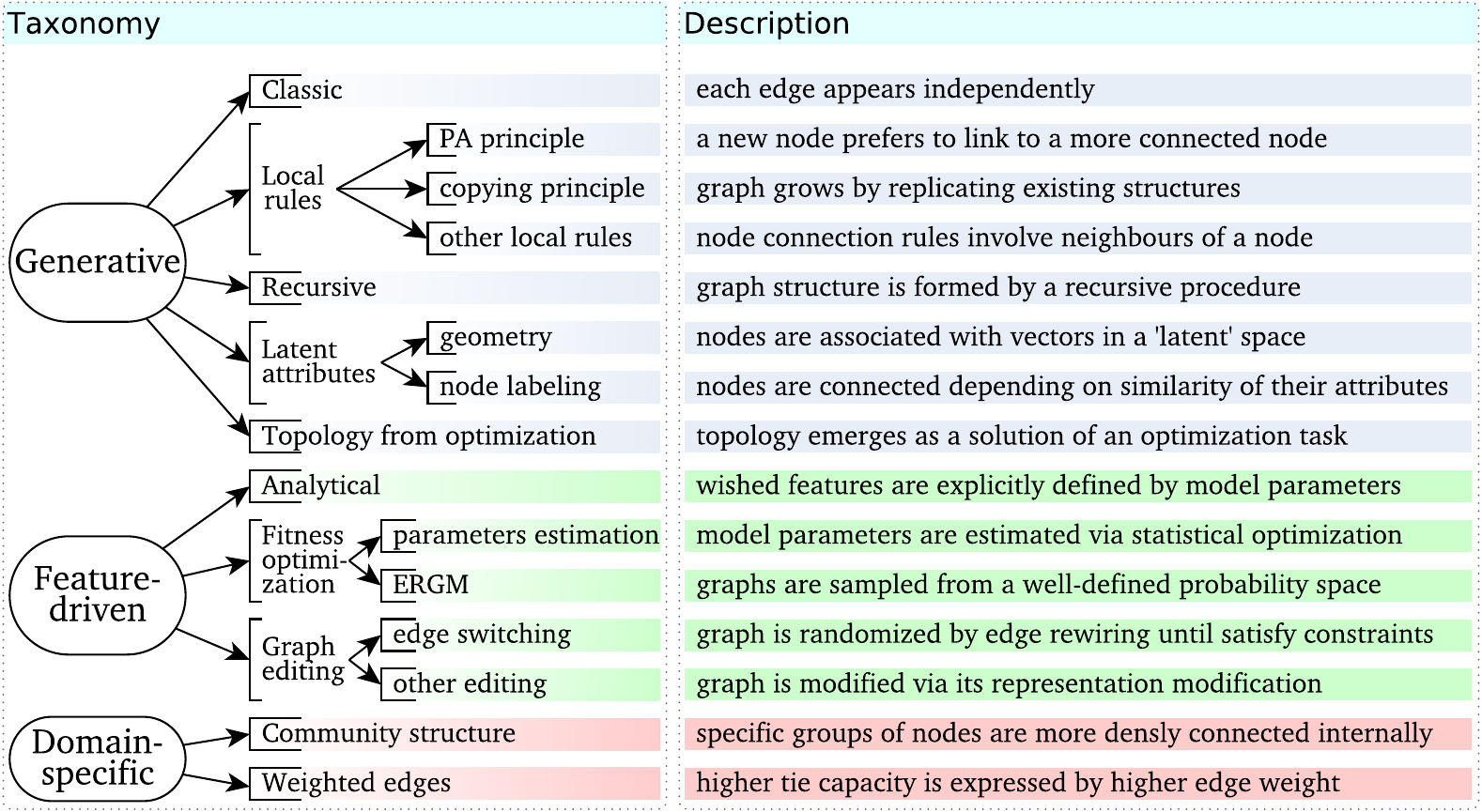}
\caption{Taxonomy of concepts, appearing in random graph modeling approaches with short descriptions of each category.}
\label{fig:taxonomy}
\end{figure}

We suggest a hierarchical taxonomy of RG concepts considering three upper-level classes based on underlying motivations (Figure \ref{fig:taxonomy}). 

\begin{enumerate}
\item \textbf{Generative} class covers all graph generating mechanisms invented to qualitatively explain graph patterns. The relevant model development order is to construct a graph according to specified rules and find out what features it has, then analyze whether its features correspond to real-world graph patterns and modify the rules accordingly.
\item \textbf{Feature-driven} class focuses on designing a model, which quantitatively fit the required graph features. The development order is the opposite: given a set of desired features, one tries to design or tune a model, satisfying these features.
\item \textbf{Domain-specific} class concerns methods for generating graphs with additional network attributes, such as community structure or edge weights.
\end{enumerate}

First two classes are intended to cover all models for simple and directed graphs, while Domain-specific class covers other types of graphs which are potentially unlimited. Each class contains several categories reflecting distinctive directions of thought. Coarse-grained categories are divided into subcategories. We describe and analyze them below in details and illustrate them with particular models. Naturally, several models appear in several categories since they employ several concepts. Although these categories do not refer to all relevant models, our goal is to illustrate the concepts that cover the majority of famous RG models and generators.

\subsection{Generative class}

Starting from the simplest ER model, which is the most general and, at the same time, the least realistic model of a random graph, designers of RG models developed many algorithms trying to explain patterns presented in real networks. Barab\'{a}si-Albert model exhibits power law degree distribution as a result of preferential attachment principle. Wattz-Strogats model achieves low diameter, for so-called "small-world" networks, by random wiring in a regular lattice. Forest Fire model shows densification law and shrinking diameter pattern in evolving graphs using a recursive process, resembling forest fires, and so on. Further work in this area is an adaptation of original concepts to directed edges, introducing new heuristics, trying to combine various features in one model, etc. While such works do not suggest new concepts, we do not mention them.

The concepts comprised here represent the whole range of random graph generating approaches we are aware. We group them into five categories: \cat{Classic}, \cat{Local rules}, \cat{Recursion}, \cat{Latent attributes}, and \cat{Topology from optimization}.

\subsubsection{Classic}

\begin{figure}[h!]
\minipage{0.3\textwidth}
  \centering
  \includegraphics[scale=0.35]{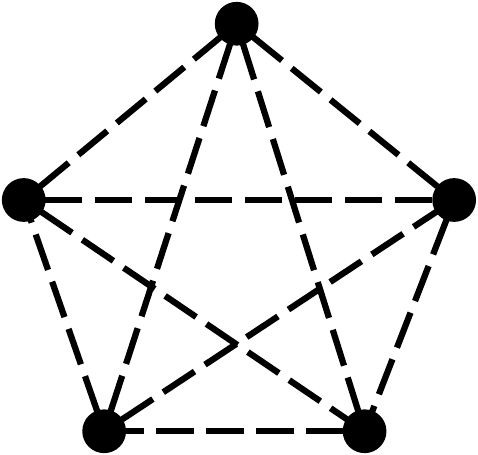}
  \caption{ER model: each edge appears with equal probability.}
  \label{fig:er}
\endminipage
\hfill
\minipage{0.68\textwidth}
  \includegraphics[scale=0.25]{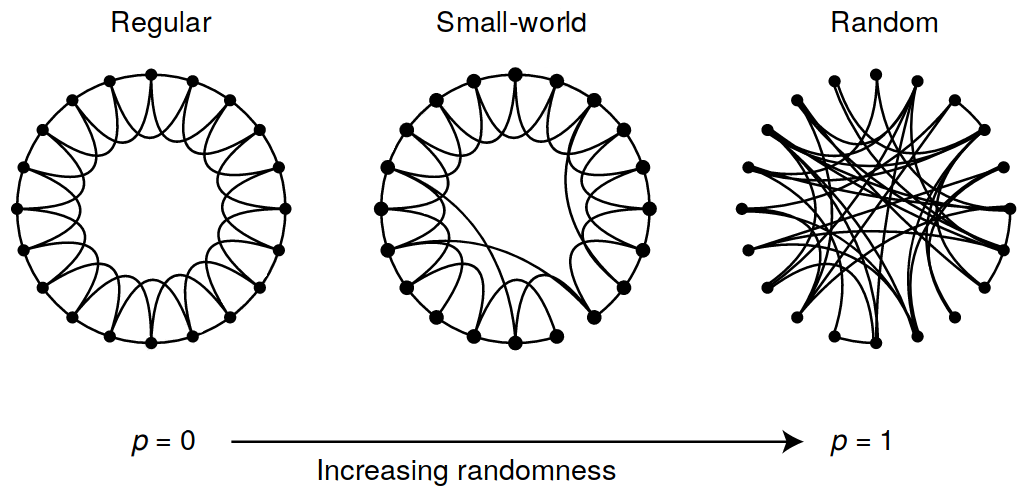}
  \caption{Intermediate values of the ratio $p$ of randomly rewired edges correspond to a small-world region~\cite{watts1998collective}.}
  \label{fig:small-world}
\endminipage
\end{figure}

The naive interpretation of randomness is to connect each pair of nodes independently. One of the first such models, the ER model~\cite{erdds1959random}, became classic: on a set of $n$ nodes, each edge appears with a constant probability $p$ (Figure~\ref{fig:er}). Although the ER model has unrealistic properties (Poissonian DD, very low clustering, etc.), it is rich with theoretical results, e.g., phase transition theory~\cite{bollobas2001random}.

Another prominent construction, named the small-world model, is aimed to achieve the low diameter together with high clustering. Watts-Strogatz model~\cite{watts1998collective} starts with a regular lattice, where each node has $k$ neighbors. Each edge is then replaced with a random edge with probability $p$. There exist intermediate values of $p$ between $0$ (regular lattice) and $1$ (ER graph) corresponding to a "small-world" region where clustering is still high and average path length decayed (Figure~\ref{fig:small-world}).


\subsubsection{Local rules}

Following A.\,V\'{a}zquez~\cite{vazquez2003growing}, by "local," we mean that the graph growth process is guided by rules involving a node with its neighbors.
One of such rules, motivated by an observation very popular in the social domain, is called "triadic closure"~\cite{granovetter1977strength}. It says that the probability of edge $(u, v)$ is higher given that nodes $u$ and $v$ have a common neighbor. This rule is expressed as a high clustering coefficient in real graphs comparing to an independent connecting of nodes in the ER model.

\paragraph{Preferential Attachment principle}

\begin{figure}[ht!]
\hfill
\minipage{0.45\textwidth}
  \centering
  \includegraphics[scale=0.4]{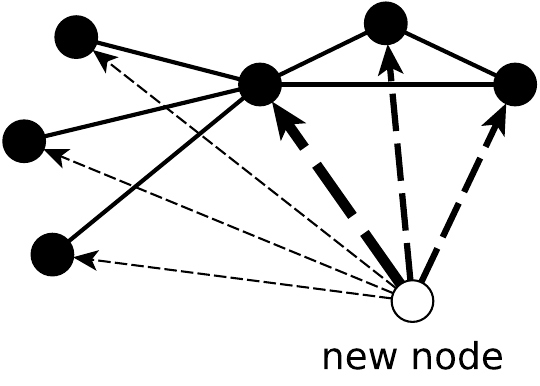}
  \caption{PA: a new node more likely connects to a more connected node. Dashed edge thickness correspond to linking probability.}
  \label{fig:pa}
\endminipage
\hfill
\minipage{0.45\textwidth}
  \centering
  \includegraphics[scale=0.4]{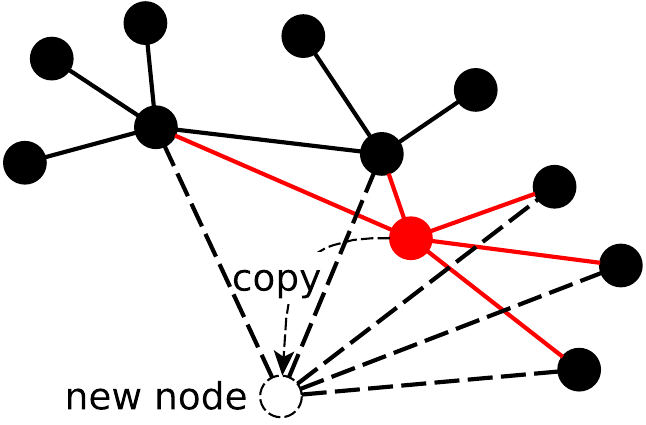}
  \caption{Copying principle: an existing part of the graph is copied, e.g., a node with its edges.}
  \label{fig:copying}
\endminipage
\hfill
\end{figure}

Two factors: the growth of the graph and the idea of linking a new node more likely to a more connected node --- together lead naturally to the power law DD. In this way, PA is employed in the Barab{\'a}si-Albert model~\cite{barabasi1999emergence} to explain scale-free property observed in many real-world networks. PA principle is vastly used in RG models, therefore a lot of variations exist. Original formulae states edge probability to be proportional to node degree: $P \sim d_i$, normalized over all nodes $i$ already presented in the graph (Figure~\ref{fig:pa}). But this predetermines a power law exponent $\gamma=3$~\cite{barabasi1999emergence}. Most notable evolution steps of PA include the following.

\begin{itemize}
\item Introduction of new parameters to PA rule, e.g., $P = \frac{A+d_i}{\sum_i{(A+d_i)}}$ allows flexible power law exponent $\gamma = 2+\frac{A}{\Delta m} \in [2, \infty)$, where $\Delta m$ is a number of new edges to be added at each step, $A$ is an extra parameter~\cite{dorogovtsev2003evolution}.

\item Modification of PA rules. In Bollob{\'a}s-Riordan model~\cite{bollobas2003mathematical}, a graph $G_1^{n}$ with $n$ nodes and $1 \cdot n$ edges is built first. $G_1^{n}$ is constructed from $G_1^{n-1}$ by adding 1 node with 1 edge according to PA rule. To obtain $G_k^{n}$ with $n$ nodes and $kn$ edges, one builds $G_1^{kn}$, split its $kn$ nodes into $k$-node groups, and collapse them, preserving the edges (edges within one group become self-loops). One of the results is that the diameter is $\approx \dfrac{\ln n}{\ln \ln n}$, which fits to the empirical value 6 for the Internet in 1999.

\item Nonlinear PA. One may generalize PA rule, linear from node degree, to an arbitrary function. For instance, $P \sim (1+d_i)^\beta - \lambda$, where parameters $\beta, \lambda$ are to be fitted: for real networks best $\beta$ varies from 0 to 1.6~\cite{kunegis2013preferential}.
\end{itemize}

PA serves as a basis for a lot of later models, which also introduce community structure, higher clustering~\cite{toivonen2006model}, and so on.

\paragraph{Copying principle}

Quite a natural mechanism of networks formation is duplicating of its parts, possibly with mutations (Figure~\ref{fig:copying}). Patterns copying takes place in various real networks. Genes can duplicate during the evolution process. Thus their interaction edges are duplicated in protein interaction networks. In WWW as well as in citation networks, authors could inherit most links from one page (work) to another on a similar topic.

Original formalization by Jon M.\, Kleinberg et al.~\cite{kleinberg1999web} includes the four processes acting at each iteration: node creation/deletion and edge creation/deletion with some probabilities. The essence of the model is the edge creation process. A node $v$ to add edges for, and the number of edges $k$ to be added are sampled from predefined distributions. With probability $\beta$, node $v$ is linked to $k$ randomly chosen nodes, and with probability $(1-\beta)$, edges of a randomly chosen node $u$ are copied.
Such a copying model produces the power law DD with $\gamma=\frac{2-\alpha}{1-\alpha} \in [2;3]$ depending on the growth factor $\alpha = \frac{\beta}{1-\beta}$.
It is also shown to demonstrate a large number of bipartite cliques (as in the Web graph), creating some community effect~\cite{kumar2000stochastic}.

In a Growing network model with copying~\cite{krapivsky2005network}, in addition to copying edges of a target node $u$, a chosen node $v$ also connects to $u$ itself. This provides that the number of edges $m$ grows faster than the number of nodes $n$, which was observed in real world networks as densification law.

A kind of mutations could be introduced, like in Duplication divergence model~\cite{vazquez2003growing}. Here, after copying edges for each of the neighbors $j$, one of the two edges $(u, j)$ or $(v, j)$ are removed with probability $1-q_e$. Notably, the clustering coefficient as a function of node degree shows power law decay with exponent depending on $q_e$.

Algorithm for replicating of complex networks (ReCoN)~\cite{staudt2016generating} copies a given graph $k$ times and then applies edge switching to make the replicas connected and add randomization. Although simple, ReCoN is shown to preserve the Gini coefficient of the DD, relatively high clustering coefficient\footnote{Due to separate edge switching within communities and between communities, CC does not fall much. Generally speaking, edge switching breaks clustering features.}, and small diameter.


The concept of structure copying is present in many RG models often implicitly or among other mechanisms. For instance, in the Forest Fire model~\cite{leskovec2007graph}, a new node attaches to the neighbors of its target node (with "burning" probability) and this "burning" process continues recursively. The GScaler algorithm~\cite{zhang2016gscaler} decomposes the input graph into separate nodes with edge stubs, multiplies them, and rewires according to the edge correlation function.

\paragraph{Other local rules}

In the world of graph growth models, perhaps as a further evolution of PA principle, various local based approaches emerged. They were shown to explain other important features like degree correlations and an inverse proportionality between the clustering coefficient and the vertex degree~\cite{vazquez2003growing}.
Now we give examples of different local rules employed in models.

Random Walks model~\cite{vazquez2003growing}. A new node $v$ connects to a randomly chosen existing node $w$. Then, with some probability $q_e$ it connects to one of its neighbors $w'$. If an edge is created, proceed to a neighbor of $w'$ and so on, thus performing a random walk.
As a modification, node $v$ could try to connect to each of $w$'s neighbors, which resembles an exhaustive search.
These random walk rules lead to the power law in-DD and relatively high clustering.

Nearest Neighbors model~\cite{vazquez2003growing}. A new node $v$ also connects to $w$, and then with probability $p$ it connects to one of its neighbors. Besides power law DD, this simple mechanism provides two non-trivial patterns, observed in social networks. Clustering coefficient as a function of node degree follows power law; average neighbor degree increases as a function of node degree.

Forest Fire model~\cite{leskovec2007graph}. The first step is the same: a new node $v$ connects to $w$. Among its unvisited neighbors, it selects $x$ ones, reachable via out-links and $y$ ones, reachable via in-links (or as much as possible, if not enough). Node $v$ creates out-links to the selected nodes, marking them as visited, and the process continues recursively. $x$ and $y$ are sampled from geometric distributions parameterized with forward $p$ and backward $rp$ burning probabilities.
Surprisingly, this model demonstrates a set of significant features: heavy-tailed in- and out-DD, densification power law, and shrinking diameter. According to the experiments with social networks, Forest Fire model also shows the clustering coefficient consistency with real data~\cite{sala2010measurement}.

The most popular local based heuristic involves creation of triadic closures. They could be formed with some probability at each iteration of an algorithm. For example, two random neighbors of node $i$ are linked if are not already~\cite{davidsen2002emergence}, or friend of friend of node $i$ is linked to $i$~\cite{marsili2004rise}. These models also exploit random node deletion (with some probability at each step)~\cite{davidsen2002emergence} or random edge deletion~\cite{marsili2004rise}, therefore, a permanent growth becomes a dynamical evolution. The process continues until stationary distributions (DD, average degree) is reached.


\subsubsection{Recursion}

One of the substantial insights into the structure of complex networks concerns their self-similar nature. Nodes in social networks, as well as in computer networks, form communities (more tightly connected groups), consisting of smaller communities, and so on. Another indication of hierarchical organization is scale-free property, together with high clustering~\cite{ravasz2003hierarchical}. Therefore, a set of RG models, grounded on recursive algorithms, were suggested.

\begin{figure}[h!]
\minipage{0.5\textwidth}
  \includegraphics[scale=0.14]{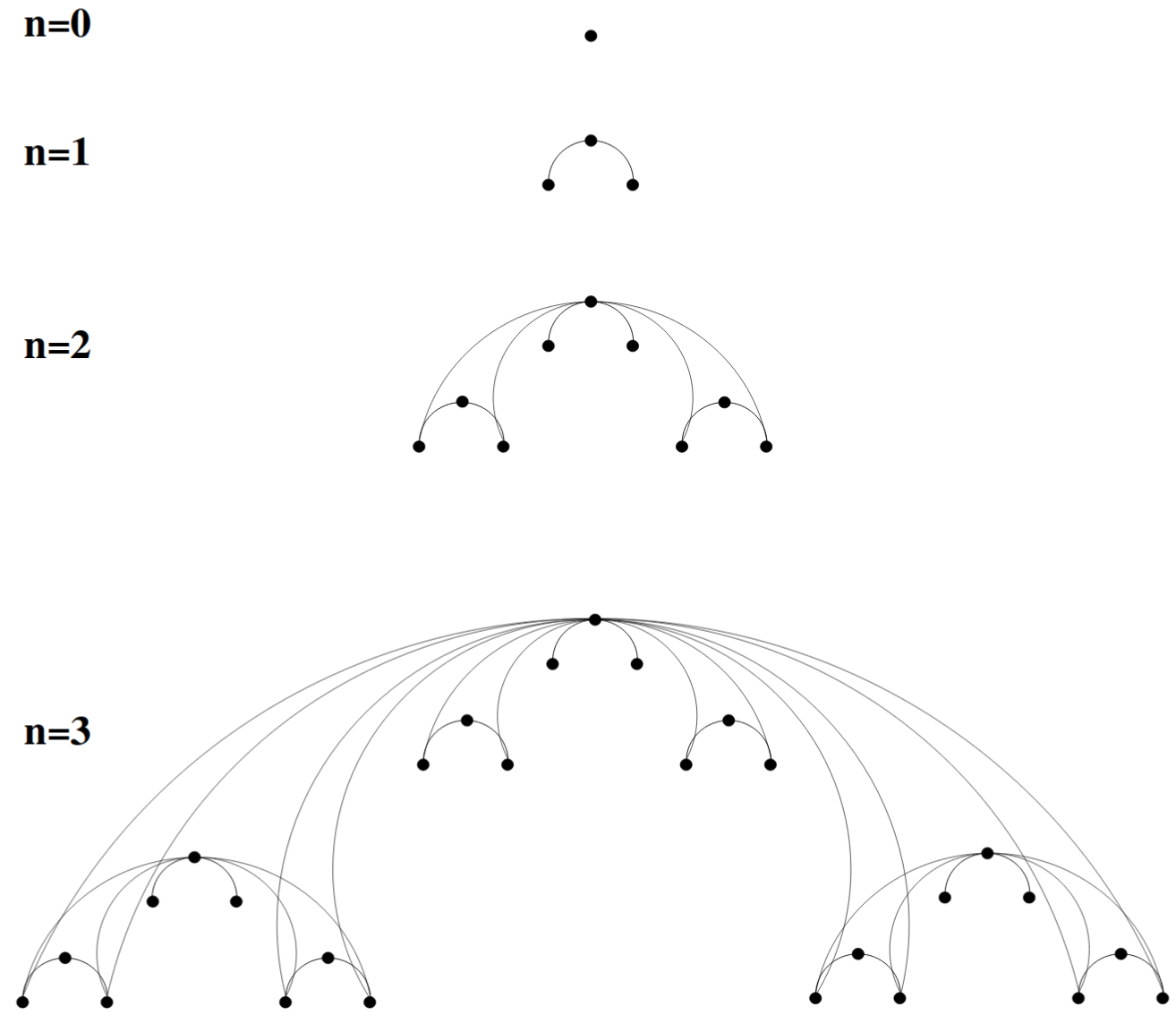}
  \caption{Deterministic recursive graph construction~\cite{barabasi2001deterministic}.}
  \label{fig:recursive_determ}
\endminipage
\hfill
\minipage{0.48\textwidth}
  \includegraphics[scale=0.35]{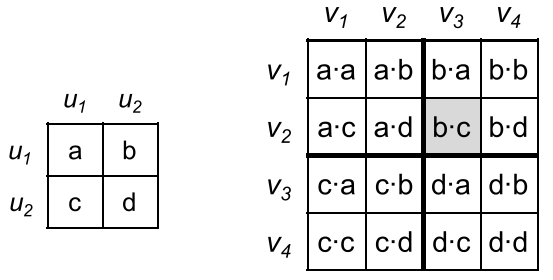}
  \caption{Recursively defined probabilistic adjacency graph matrix. For instance, the edge $(2,3)$ sampling probability equals $b \cdot c$~\cite{leskovec2010kronecker}.}
  \label{fig:recursive_matrix}
\endminipage
\end{figure}

A straightforward deterministic method starts with a small initial graph $G_0$ (or one root node) and at each step creates $N$ replicas of the current graph $G_k$ (Figure~\ref{fig:recursive_determ}). The replicas are linked with each other in the same manner as $G_0$, e.g., the root node links to all nodes at the bottom level~\cite{barabasi2001deterministic}. It is proved that a deterministic recursive procedure gives power law DD, high clustering, and CC inverse to node degree $CC(d) \sim d^{-1}$~\cite{dorogovtsev2002pseudofractal}.
Other variants are based on iterative addition of $a \cdot d_i$ new nodes to each of the existing nodes $i$, combined with edge rewiring~\cite{molontay2015fractal}. Replacing each edge with two parallel paths, consisting of $u$ and $v$ links ($(u,v)$-flower)~\cite{rozenfeld2007fractal}. And, finally, replacing each edge with the initial graph~\cite{xi2017fractality}.

One of the most influential concepts, based on a recursively defined matrix, appears in R-MAT algorithm~\cite{chakrabarti2004r}. Graph adjacency matrix $A_{ij}$ of size $2^k \times 2^k$ is recursively partitioned into four equal quarters until reaching one cell. The graph has $n=2^k$ nodes. Its edges are sampled with the help of these partitions according to the four defined probabilities $a, b, c, d$ of getting into each quarter. To sample an edge, a selection of the quarter, then subquarter, and so on is made $k$ times, resulting to a particular cell $(i,j)$.

In another interpretation by J.\,Leskovec, D.\,Chakrabarti, J.\,Kleinberg, and C.\,Faloutsos~\cite{leskovec2005realistic}, an initially defined probability matrix $A_{ij}^1$ is multiplied by Kronecker multiplication $k$ times, resulting in a probabilistic adjacency matrix $A_{ij}^k = (A_{ij}^1)^{\otimes k}$.
Along with being mathematically parsimonious, this recursive procedure provides a set of useful graph features, if one appropriately specifies the initial parameters. Namely, multinomial in- and out-DD, low diameter, multinomial eigenvalue and eigenvector distributions, and hierarchical community structure. In the embodiment, called Stochastic Kronecker Graphs (SKG)~\cite{leskovec2010kronecker}, it is shown to demonstrate densification power law. The Kronecker multiplication is crucial here, since Cartesian product of graphs or aforementioned construction~\cite{barabasi2001deterministic} do not yield densification power law graphs.

The SKG model is deeply studied and rich of extensions due to its mathematical tractability, low generation complexity, and additional procedure of parameters fitting~\cite{leskovec2010kronecker}. The extensions include adding a random noise to overcome DD oscillating~\cite{seshadhri2011depth}; introducing tied parameters to increase graphs variability for domain imitating~\cite{morenomodeling}; introducing multiple fractal structures in the model to expand space of covered graphs~\cite{moreno2013block}.

A closely related concept underlies Multi-fractal network generator (MFNG)~\cite{palla2010multifractal}. In addition to the recursively specified edge probability ($P_{ij} = \prod_{q=1}^k{p_{i_q} p_{j_q}}$ with $l$ probabilities $p_{i_q}$ as parameters), nodes belong to recursively defined categories. Namely, $[0,1]$ interval is split into $l$ different subintervals defined by extra $l-1$ parameters. Each of the intervals is iteratively split again with the same ratios $k$ times, thus defining the categories. Graph nodes are uniformly sampled as points in $[0,1]$. This procedure gives a more flexible model which is supplied with a fitting procedure.

The concept of recursive topology construction is well consistent with the fractal structure of real networks. It also explains a set of power laws (DD, CC vs. node degree, eigenvalues) and the low diameter. However, recursion-based algorithms often generate graphs with $n=n_0^k$ nodes, which could be too coarse-grained for practical purposes.


\subsubsection{Latent attributes}

The idea is to assume that linking probability depends on some inherent properties of the nodes expressed as their attributes.
Motivation from the social domain is called homophily, which claims that similarities attract: people of close age, interests, occupation, geographical location, etc. are more likely to be connected within the network~\cite{mcpherson2001birds}.
This concept is formalized via incorporating node attributes in the model and stating edge probability as a function of node attributes: $P_{ij} = f(\vec{a}_i, \vec{a}_j)$. Such models are also referred as "spatial" or "latent space", meaning attributed nodes as points in a space of social attributes.

This category of concepts we divide into two directions: geometry and node labeling.

\paragraph{Geometry}

An intuitive interpretation of nodes' attributes as geographical coordinates is productive in modeling ad hoc wireless networks, sensor-actuator networks, and the Internet, where physical distance between the nodes directly influences their connectivity~\cite{onat2008generating}.

\begin{figure}[h!]
\minipage{0.48\textwidth}
  \includegraphics[scale=0.32]{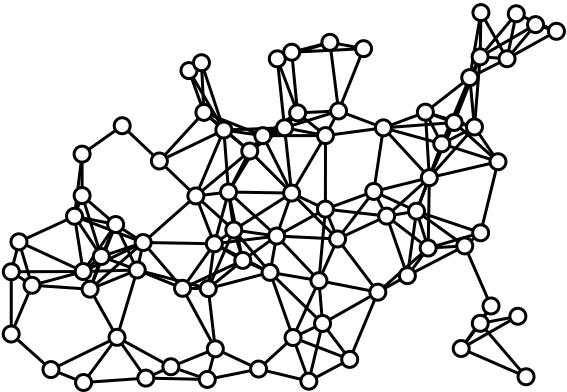}
  \caption{Random geometric graph. Nodes are points, randomly distributed in a space. Edge probability depends on the nodes' coordinates.}
  \label{fig:geom}
\endminipage
\hfill
\minipage{0.48\textwidth}
  \includegraphics[scale=0.14]{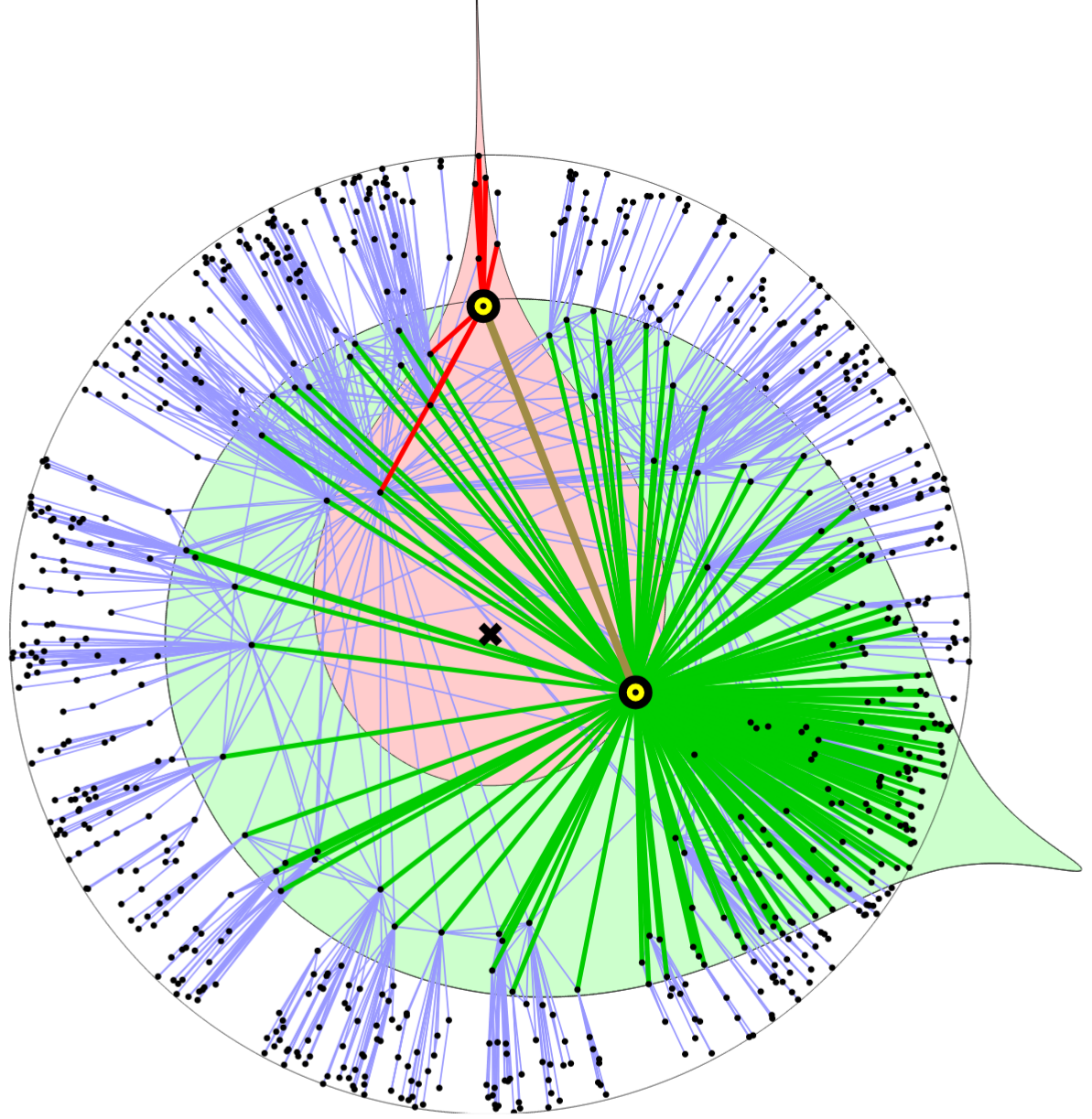}
  \caption{Hyperbolic random graph: edge probability depends on the hyperbolic distance between nodes. Nodes are distributed within a hyperbolic disk of radius $R$. Green and red areas correspond to hyperbolic disks of radius $R$ centered at the highlighted nodes~\cite{krioukov2010hyperbolic}.}
  \label{fig:hyperb}
\endminipage
\end{figure}

Common approaches follow this scheme. First, $n$ points are distributed in 1 or 2-dimensional area in Euclidean space, usually uniform in $[0;1]^2$, or a Poissonian point process is used. Then, edges are sampled probabilistically according to the distance between nodes $dist(i,j)$ (Figure~\ref{fig:geom}). The dependency function varies across the works: 
exponential decay $P_{ij} \sim e^{-\alpha dist(i,j)}$ in Waxman model~\cite{waxman1988routing};
power decay $P \sim \frac{{d_i}^{\alpha}}{{dist(i,j)}^{\sigma}}$ in S.-H.\,Yook et al.~\cite{yook2002modeling} with best fit $\alpha=\sigma=1$ to the Internet;
step function $P_{ij}=p_a$, if $dist(i,j) < H$, else $P_{ij}=p_b$~\cite{wong2006spatial}.
Specifying a distribution of node points as a mixture of distributions, naturally models a community structure, e.g., a sum of Multivariate normal distributions is used~\cite{handcock2007model}.

Although achieving good results at CC, degree correlations, and community structure in these models, random geometric graphs have Poissonian DD~\cite{penrose2003random}. The remedy could go from static to dynamic model employing PA principle as the BRITE generator: $P_{ij} \sim d_j \cdot e^{-\alpha dist(i,j)}$~\cite{medina2000origin}.

If we change the distance between nodes to a cosine similarity of their vectors, we come to dot-product graphs. The nodes reside in a multidimensional space. The edge probability is given as a function of a dot-product of their vector representations: $P_{ij} = f(\vec{r}_i \cdot \vec{r}_j)$~\cite{nickel2008random}.
In a generative model, vectors $\vec{u}$ and $\vec{v}$ are sampled independently for each node from probability distributions $U$, $V$ respectively, namely, $\mathcal{U}^{\alpha}[0,1]$ --- $\alpha$-th power of uniform distribution. Corresponding nodes are connected with probability $P_{ij} = \vec{u}_i \cdot \vec{v}_j$. Together with the sparse case of $P_{ij} = \frac{\vec{u}_i \cdot \vec{v}_j}{n^b}$, $b \in (0, \infty)$, the model is thoroughly studied theoretically and shown to generate power-law graphs with small diameter and high clustering coefficient~\cite{nickel2008random}.
Node vector could be interpreted as a list of interests of a corresponding individual in a modeled social network (users with common interests are more likely to communicate), or as topics of a corresponding website (related websites are more likely to be linked).

The attempts to adapt complex networks for geometric framework led to the assumption that hyperbolic geometry underlies their structure. It was shown that DD heterogeneity and strong clustering reflect the hyperbolic nature underneath~\cite{krioukov2010hyperbolic}. For example, power law exponent is a function of the space curvature. In other words, a more relevant distance metric on graphs is based on the shortest path (geodesic line), and it is rather hyperbolic than Euclidean. Moreover, hierarchical structure and tree-like patterns, common in real networks, better fit into hyperbolic space.

The standard model of Hyperbolic Random Graph utilizes a hyperbolic disk of radius $R=2\log{n}+C$. $n$ nodes are randomly distributed points with radial density $p(r) = \alpha \frac{\sinh(\alpha r)}{\cosh(\alpha R)-1}$ and uniform by angle. Pairs of nodes with the hyperbolic distance less than $R$ are connected (Figure~\ref{fig:hyperb}). In this setting, the DD is proved to be power law with exponent $2\alpha+1$, CC is non-vanishing as $n \rightarrow \infty$~\cite{gugelmann2012random}, the size of the second largest component is $O(polylog(n))$~\cite{kiwi2015bound}, and established are bounds on the diameter~\cite{friedrich2018diameter}.

A model called Geometric Inhomogeneous Random Graphs (GIRG) is claimed to (almost surely) contain Hyperbolic Random Graph as a subclass and to be technically simpler~\cite{bringmann2016geometric}. It mixes Chung Lu and geometric approaches. Nodes are randomly distributed points $\vec{x}_i$ in a $d$-dimensional torus with Euclidean distance. Like in Chung Lu model node weights $w_i$ are defined corresponding to the expected degrees. The edge probability combines geometric and Chung Lu components: $P_{ij} = \Theta (\min \left\lbrace \dfrac{1}{|| \vec{x}_i-\vec{x}_j || ^ {\alpha d}} \cdot \left( \dfrac{w_i w_j}{W} \right) ^{\alpha}, 1 \right\rbrace)$. With appropriate parameters values, a set of properties is proved to hold for GIRG: power-law DD, high CC, presence of a unique giant connected component, poly-logarithmic diameter, and small separating sets; average path length is of order $O(\log \log n)$~\cite{keusch2018geometric}.

In Embedding based random graph model (ERGG)~\cite{drobyshevskiy2017learning}, each node of a directed graph is associated with a vector $\vec{r}_i$ being a triple $\vec{u}_i, \vec{v}_i$, and $Z_i$. Link probability is based on a directed softmax model, where the conditional probability of the edge $i \rightarrow j$ is: $P(j|i) = \exp (\vec{u}_i \cdot \vec{v}_j - Z_i)$, with $Z_i$ being a normalization coefficient~\cite{ivanov2015learning}. At the construction phase, edge $i \rightarrow j$ is created iff $P(j|i)$ is above a threshold $t_G$. Representations $\{\vec{r}_i\}$ and the threshold are learned to fit best to a given graph $G$.

As a resume, we note that the selection of graph geometry could be treated as the selection of metric in the node vectors space. The simplest geometry is Euclidean one. Dot-product based metric reflects spherical geometry (due to cosine similarity). More sophisticated and efficient is the hyperbolic metric.

\paragraph{Node labeling}

Besides geometric interpretation, the concept of representing the node as a vector of attributes takes another form. The key assumption is that edge probability defined by the similarity of node labels.

In Random typing graphs (RTG)~\cite{akoglu2009rtg}, a random typing process is used to generate character sequences terminating with "space". Each unique word corresponds to a node. At each algorithm step, source and destination node labels are created in parallel by one letter $l$, each having its own typing probability $p_l$. An edge is created between the nodes or edge weight is incremented if it exists. Additionally, in order to model the homophily (and community structure), an imbalance factor $\beta < 1$ is introduced. $\beta$ diminishes generating the probability of different letters at the same position, i.e., $p(a, b) = \beta p_a p_b$, while $p(a, a) = p_a p_a$. This trick makes nodes with similar labels be connected more often. RTG model emerges seven power law dependencies: DD; densification; number of triangles a node participate; eigenvalues of adjacency matrix; largest eigenvalue versus the number of edges $m$; total edge weight depending on $m$; and node strength depending on its degree.

In R-MAT~\cite{chakrabarti2004r}, as well as in SKG~\cite{leskovec2010kronecker} approaches, the initial probability matrix $\mathbf{A}^1$ can be treated as individual attributes similarities. Thus, each node becomes a unique sequence of $k$ attributes, where $k$ is a value of Kronecker power. Edge probability equals to the product of these individual similarities for two nodes. In this way, higher diagonal values of $\mathbf{A}^1$ correspond to the homophily principle, since the coincidence of attributes increases edge probability.



\subsubsection{Topology from optimization}

One interesting approach concerns a concept of network topology emerging as a solution of some optimization task. One could say that organization of many biological systems, the Internet, and communication networks were formed as a result of adaptation to the environment under the constraints and maximization the network efficiency. Therefore, the network structure can be derived through optimization of a fitness function.

A Heuristically Optimized Trade-offs Model~\cite{fabrikant2002heuristically} is aimed to explain power law DD in the Internet graph as a result of locally made trade-offs. Nodes in the model are sampled uniformly in a unit square. When a new node $i$ appears, it chooses node $j$ to connect to by minimizing two goals: geographical distance to it $dist(i,j)$ and a centrality $h_j$ (e.g., the average path length from $j$ to all other nodes in the graph), i.e., $\alpha dist(i,j) + h_j \rightarrow min$. Intermediate values of parameter $\alpha$ correspond to the emergence of power law as a trade-off between geographical and centrality constraints.
This model is generalized by N.\,Berger et al.~\cite{berger2004competition}, who show that the competition between connection cost and routing cost causes PA behaviour.

Various simple topologies can emerge from the maximization of a survival fitness function:\linebreak $\alpha \eta_E + (1-\alpha) \eta_R - C \rightarrow max$~\cite{venkatasubramanian2004spontaneous}. Here $\eta_E$ reflects the efficiency of system functioning, formalized as an inverse of the average graph path length. $\eta_R$ is robustness to potential damage (such as node/edge removal), non-trivially expressed via sizes of strongly connected components after a node removal. $C$ refers to resource constraints, measuring the cost of node and edge addition. By means of simulations, there were obtained "star", "hub", "circle" and power law topologies.


\subsection{Feature-driven class}

Early graph models were aimed to qualitatively explain the main patterns, observed in the real networks. However, it is more useful not only to capture the important graph features, but to be able to control them parametrically. If a model allows custom power law exponent and clustering coefficient, it becomes a much more flexible and efficient instrument for network analysis.
Unfortunately, in practice, model parameters influence on resulting graph properties in a very complicated way. Moreover, known graph measures are not independent of each other and could not take arbitrary values.
To address this problem the RG models are often supplied with parameter estimation procedures, aimed to fit the requirements.
Model fitting algorithm is a key point of models in the Feature-driven class.

In contrast to the Generative class, the Feature-driven class concerns approaches, which whether take as input a list of features, desired to be reproduced in output graphs, or directly fit a given graph, implicitly learning its features. Many modern models combine paradigms of both classes, e.g SKG were merely a graph generator until a parameter fitting procedure Kronfit was invented.

We distinguish three categories of approaches each of which is rich of variations: \cat{analytical way}, \cat{fitness optimization}, and \cat{graph editing}.

\subsubsection{Analytical way}

Quite a straightforward approach is to design a graph generating algorithm in a way such that its parameters could be analytically found given the wished graph features.
Such a model is mathematically tractable, allows for precise control of graph features and thus useful for analysis.

Simplest cases include the realization of prescribed degree sequence, either fully custom or sampled from a family of distributions like power law or Double Pareto Log-Normal distribution~\cite{seshadri2008mobile}.

Configuration model~\cite{bender1978asymptotic} implements a sequence of node degrees $\{d_i\}_{i=1}^n$: each node is assigned with $d_i$ edge stubs which are then wired randomly.
Plenty of models grew from this concept, refer to D.\,Chakrabarti and C.\,Faloutsos~\cite{chakrabarti2006graph} for details.
In an Expected Degree model aka Chung Lu model~\cite{chung2002average, chung2002connected} each node $i$ is given with an expected degree $w_i$, edge probabilities being $P_{ij} \sim w_i w_j$.
Generalized Binomial Graph~\cite{kovalenko1971theory} defines a matrix of edge probabilities itself as a parameter: $\mathbf{P}=[P_{ij}]$.

Being quite simple, these models are well studied for various power law exponents, emergence of connected components, size of largest cliques, etc.~\cite{janson2010large}.
Although being poor models for real networks, such constructions widely serve as null-models. A class of all graphs with the same nodes degrees is a classic null model. It is used for network motif detecting task~\cite{fosdick2018configuring}.

\begin{figure}[h!]
\minipage{0.58\textwidth}
  \includegraphics[scale=0.4]{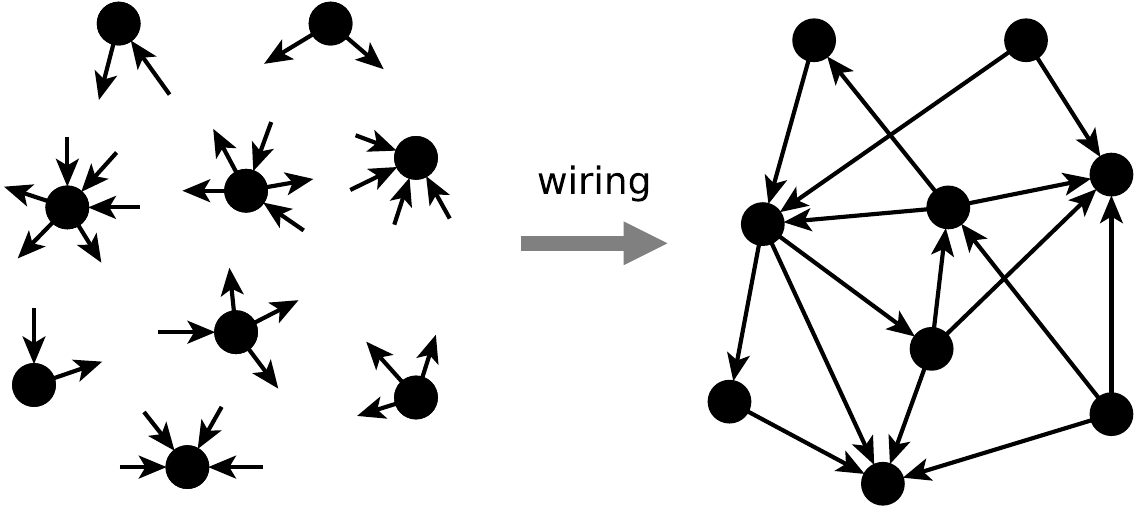}
  \caption{Configuration model: each node has edge stubs corresponding to its degree. Edge stubs are then randomly wired.}
  \label{fig:config}
\endminipage
\hfill
\minipage{0.35\textwidth}
  \centering
  \includegraphics[scale=0.48]{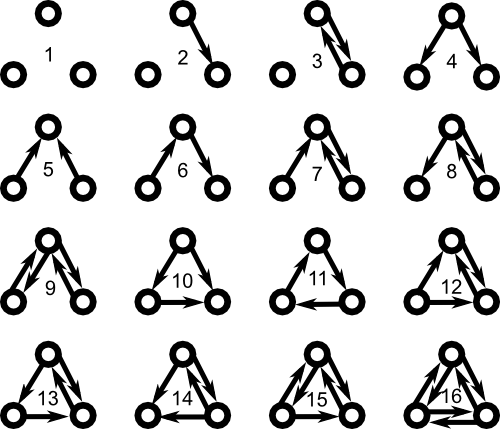}
  \caption{16 possible subgraph configurations on 3 nodes. Their exact distribution in the generated graph could be expressed analytically in a Triplet model~\cite{wegner2011random}. (Picture from \url{https://mathinsight.org/evidence_additional_structure_real_networks}.)}
  \label{fig:subgraphs}
\endminipage
\end{figure}

More complex task is to reproduce the desired subgraph distribution in a graph.
A Triplet model by A.\,Wegner~\cite{wegner2011random} considers generating one of four possible edge configurations (having from 0 to 3 edges) on each node triplet, according to the probabilities $p_1, ..., p_4$. There are 16 variants in the case of directed edges (Figure~\ref{fig:subgraphs}). Subgraph distribution in the generated graph is expressed via four (or 16) equations, which connect their probabilities to the probabilities of generating each subgraph configuration on the initial node set. A Multiplet model, generalizing to $d$-nodes subgraphs, is also described by A.\,Wegner. Unfortunately, it requires significantly more equations when $d$ increases.

Another difficulty arises when one tries to combine several features. A common method is to iteratively modify graph, consequently satisfying needed features one by one.
Implementing a target degree sequence and CC together is already non-trivial and is not solved exactly. For instance, L.\,Heath and N.\,Parikh~\cite{heath2011generating} suggest to iteratively add triangles to realize the node triangle sequence and then add single edges until degree sequence is reached. Here the resulting DD is exact while CC is close to the expected but deviates for dense graphs, presumably because tuning the DD violates CC, achieved at the first step.

Despite the absence of ways to accurately implement a set of graph features, it is often enough in practice to approximate them in exchange for the ability to control a large number of parameters.
A branch of RG generators, providing many parameters to tune, serve to construct benchmark graphs.
The most famous one could be a series of LFR algorithms~\cite{lancichinetti2008benchmark, lancichinetti2009benchmarks} for generating directed weighted graphs with overlapping community structure. LFR allows to tune in-, out-DD and community size power law exponents together with their extremal values, mixing parameter controlling the extent of communities overlapping, and others.
Such RG models usually employ simple components like ER and Configuration model, utilize greedy algorithms, and have a narrow applicability area, where parameters may be considered almost independent.

\subsubsection{Fitness optimization}

In the majority of cases, parameters of a model influence graph features non-trivially. To fit the parameters for a particular graph or to satisfy the wished feature values, a full range of methods for mathematical optimization is involved.
Traditionally, one constructs a fitness function of model parameters and optimize it, using standard techniques.

We consider 2 approaches in this category: parameters estimation and exponential models.

\paragraph{Parameters estimation}

The specificity of parameter estimation for complex networks is that the empirical data is often represented by only one graph.
A popular approach is maximum likelihood estimation, where likelihood $P(\Theta | G)$ is maximized over model parameters $\Theta$ given a graph $G$. According to the Bayesian framework, $P(\Theta | G) = P(G|\Theta) \frac{P(\Theta)}{P(G)}$ and $P(G|\theta)$ are maximized instead, assuming uniform prior $P(\Theta)$.

In SKG~\cite{leskovec2010kronecker}, an initial probabilistic adjacency matrix $\mathbf{A}^1$ must be tuned such that its Kronecker power $\mathbf{A}^k$ best fits to a given graph $G$. Power $k$ is simply a minimal one to get enough nodes.
For the rest of matrix entries $\Theta = \{\mathbf{A}^1\}_{ij}$, KronFit algorithm~\cite{leskovec2010kronecker} optimizes log-likelihood $\log{P(G|\Theta)}$ by gradient descent. The main challenge here is to take into account all possible $n!$ node permutations to match $\mathbf{A}^k$ to adjacency matrix of $G$: $P(G|\Theta) = \sum_{\sigma}{P(G|\Theta, \sigma) P(\Theta|\sigma)}$. A super-exponential summing is efficiently overcome by applying Metropolis sampling for permutations distribution $P(\sigma|G, \Theta)$, which requires $O(kn)$ steps.

In the ERGG~\cite{drobyshevskiy2017learning} model, parameters $\Theta$ consist of a triple $\vec{r}_i = \{ \vec{u}_i; \vec{v}_i; Z_i \}$ for each node and a threshold $t_G$ for edge creating. Due to the high computational complexity of direct likelihood $P(G|\Theta)$ optimization, it is replaced by its approximation $J_{\Theta}$ with the same objective. The task could be reduced to maximization of the score function $s_{ij} = \vec{u}_i \cdot \vec{v}_j - Z_i$ over all edges $i \rightarrow j$, while minimizing it over non-edge pairs. The challenge is that the space dimensionality must be low: $d \ll n$. Threshold $t_G$ is determined to best separate the edges of $G$ from non-edges according to their score $s_{ij}$.
Random graph is constructed by sampling new node vectors $\{\vec{r}'_i\}$ from the same distribution as $\{\vec{r}_i\}$, and creating edges using the computed threshold: edge $i \rightarrow j$ appears iff $s_{ij} > t_G$.

Generally, the task of mapping nodes of graph $G$ into low-dimensional vectors, encoding maximal information of $G$, is called graph representation learning or graph embedding. This direction is actively developing in recent years~\cite{goyal2018graph}. Its main benefit for RG modeling could be that it turns the graph into a set of vectors, which is much more convenient as input for machine learning algorithms.


An alternative for model parameters estimation could be the method of moments.
MFNG~\cite{palla2010multifractal} models a graph recursively, like SKG, specifying $l$ node category probabilities and $l \times l$ matrix of category similarities, but then goes in another way.
Fitting to a real graph could be done by a method of moments as a task of minimization of the deviation of a set of target features from their expected values~\cite{benson2014learning}. Strong point of this approach is that statistics, that can be formulated as events on a subset of the edges (number of edges, cliques, stars, and so on), can be analytically expressed through model parameters and thus could be used for fitting.

Since SKG model also allows to express edge-based features via model parameters, the method of moments could be applied for it~\cite{gleich2012moment}.

\paragraph{Exponential random graph models}

In a general case of the RG model, one would like to specify a graph probability space $\mathbb{P}(G)$, such that these graphs satisfy a set of wished constraints imposed on graph statistics $F(\vec{s}(G))=0$. Statistical framework suggests to choose the distribution with maximal Shannon entropy $S[\mathbb{P}]= -\sum_{G \in \mathcal{G}}{P(G)\log P(G)}$, since it gives no additional information except that contained in the constraints. Maximizing the entropy subject to constraints $\sum_{G \in \mathcal{G}}{P(G)\vec{s}(G) = \vec{s}(G^*)}$ gives exponential solutions: $P(G) = \frac{1}{Z(\vec{\theta})}e^{\vec{\theta} \vec{s}(G)}$, where the model parameters $\vec{\theta}$ are defined from constraints equations and $Z(\vec{\theta})$ is a normalization coefficient.
The idea of ERGM is to explain the observed graph $G^*$ by the statistics of its topology and node attributes.
Statistics $s_1(G), s_2(G), ...$ could be any measurable variables of network structure: number of edges, triangles, k-stars, degree sequence, or attributes: age, proximity, gender, etc.

Most ERGMs, except for trivial examples, can not be solved analytically. Exact calculation of a partition function $Z(\vec{\theta})$ and its derivatives is impossible. Therefore, approximate solutions and maximum likelihood or pseudo-likelihood methods for parameter estimation were developed~\cite{van2009framework}. Markov chain Monte-Carlo sampling is a widely employed technique, where one builds a Markov chain with a target stationary distribution.

The simplest ERGM instances are the ER model ($s(G) = m$) and Configuration model ($\vec{s}(G) = \{d_i\}$).
A more interesting well-known ERGM example is Stochastic Block Model (SBM). In SBM, graph nodes belong to one of the $Q$ groups (communities) with prior probabilities $p_i$. Nodes are linked according to an affinity matrix $\mathbf{P}$ of size $Q \times Q$, specifying inter-group probabilities. Given a real network, the model parameters could be estimated using expectation-maximization algorithm~\cite{decelle2011asymptotic}.

The key advantage of ERGMs is their probabilistic rigor, which means that the defined distribution is the best choice under given constraints $\vec{s}(G)$ in a statistical sense. This makes them attractive null-models, widely used to analyze social and biological networks.
However, at the time serious problems arise, when graphs become larger ($n > 10^4$) or conditions become more complicated than linear functions of $A_{ij}$. Computational complexity and parameters sensitivity are the main issues. A balance is needed between accuracy and speed~\cite{an2016fitting}.

\subsubsection{Graph editing}

An alternative to constructing a random graph with desired features from scratch is to randomize an existing graph, preserving its features of interest. Simplest graph editing operations include node/edge addition/removal. Many techniques are based on combinations of them.
A secondary goal of graph randomization is to introduce variability in the model. 

\paragraph{Edge switching}

\begin{figure}[h!]
\minipage{0.3\textwidth}
  \includegraphics[scale=0.35]{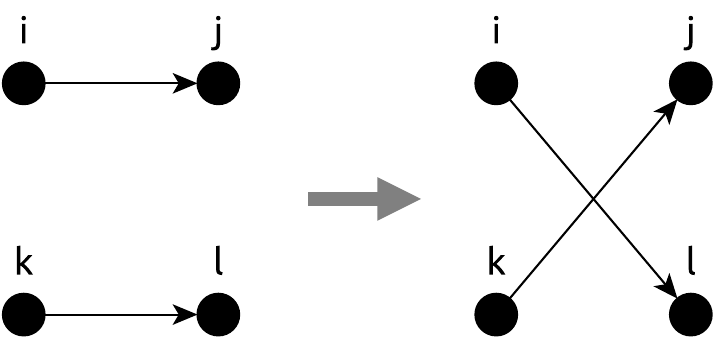}
  \caption{Edge switch operation modifies edges configuration while keeps node degree unchanged.}
  \label{fig:es}
\endminipage
\hfill
\minipage{0.6\textwidth}
  \includegraphics[scale=0.35]{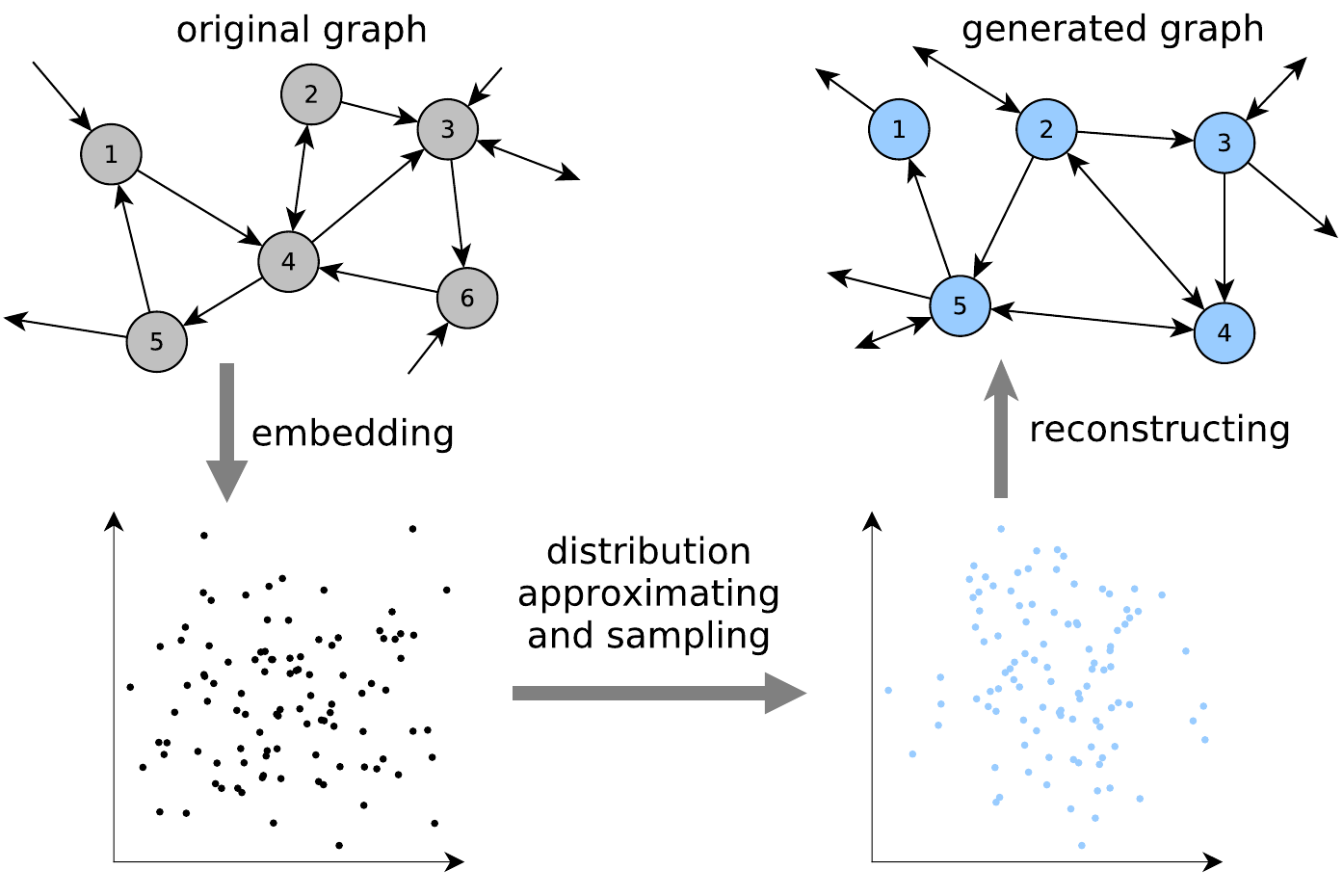}
  \caption{ERGG: a graph is modified at the level of its vector representation.}
  \label{fig:ergg}
\endminipage
\end{figure}

The classical procedure of graph randomizing is edge switching, or edge rewiring. It is repeatedly applied to modify a graph $G_C$ such that a set of constraints $C$ remains satisfied.
The most widespread operation is pairwise edge switch, since it keeps node degrees unchanged: a pair of edges $i \rightarrow j, k \rightarrow l$ is rewired into $i \rightarrow l, k \rightarrow j$ (Figure~\ref{fig:es}).

An important fact, used in many approaches, is the following.
Let a Markov chain start with an initial graph $G^0$ and a pair of edges to be switched is picked randomly at each step. Then the chain has a stationary distribution uniform over all graphs with the same node degrees. Moreover, it is irreducible, i.e., any configuration is reachable from any other. These properties make it easy to uniformly generate random graphs with given DD~\cite{taylor1981constrained}. In practice, for graph generating one waits some time, linear to the number of edges $m$, while the chain converges. Empirically, $100m$ steps is enough~\cite{milo2003uniform}.

In a case of more elaborate constraints $C$, a standard Monte Carlo sampling techniques are employed to achieve Markov chain with a wished stationary distribution corresponding to $C$. For instance, Ying Xiaowei and Wu Xintao~\cite{ying2009graph} use the Metropolis-Hastings algorithm to sample graphs with a target distribution of features $g(S)$. Namely, at the step $t$, a potential edge switch is accepted with probability $P_{G^{t-1} \rightarrow G^t} = \min{\left( 1, \dfrac{g(S(G^t))}{g(S(G^{t-1}))} \dfrac{f(S(G^{t-1}))}{f(S(G^t))}\right)}$, where $f(S)$ is the distribution of feature $S$ over all graphs with the same degree sequence.
A particular example is the ClustRNet algorithm~\cite{bansal2009exploring}, where, besides the DD, the only constraint is the CC and the graph connectivity. Thus, the transition probability is simply 1, only if the CC of $G^t$ is higher than some threshold and $G^t$ is connected, and 0 otherwise.
Similarly, one can generate $dK$-random graphs, where $C$ is $dK$-distributions~\cite{mahadevan2006systematic}.

Unfortunately, MCMC, guided by complex constraints, suffer from two problems: not all states, satisfying the constraints $C$, could be reachable from each other via allowed switches (non-ergodicity property), and an increase of chain convergence time.

To make the state space more connected, L.\,Tabourier, C.\,Roth, and J.-Ph.\,Cointet~\cite{tabourier2011generating} suggest $k$-edge switches. They are defined for $k$ edges $\{a_i \rightarrow b_i\}_{i=1..k}$, not necessarily distinct. Edges' endpoints $\{b_i\}$ are randomly permuted, resulting in $\{a_i \rightarrow \sigma (b_i)\}_{i=1..k}$ with $\sigma$ being one of $k!$ possible permutations.

Pairwise edge switch is often used as an additional randomization step in RG generators. In the ReCoN~\cite{staudt2016generating} model large graphs are generated by copying an original one (together with labelled communities) $k$ times and rewiring edges within new communities' replicas and then between them. Although edge switching preserves node degree properties, it breaks the other features\footnote{This is the consequence of the aforementioned ergodicity property. Edge switching makes reachable all possible graphs with the same degree sequence. Means that any other graph metrics can take arbitrary values. However, it depends on how a fixed degree sequence determines the other graph features.}.
In Lancichinetti-Fortunato-Radicci benchmark (LFR)~\cite{lancichinetti2009benchmarks}, edge switches are employed to adjust topological parameters: to decrease the number of intra-community edges, leaving the node degrees fixed.

\paragraph{Other editing}

Instead of modifying graph $G$ itself, one could modify its representation $R(G)$, if it properly reflects the graph features.
The problem shifts to finding an appropriate representation and convenient operations of transformation $G$ to $R(G)$ and backwards.

Multiscale Network Generation (MUSKETEER)~\cite{gutfraind2015multiscale} model suggests to use a series of coarsening-uncoarsening operations on graph Laplacian matrix $\mathbf{L}$, together with editing the coarsened state of the graph. Starting with an initial graph $G$, a sequence of repeatedly coarsened graphs $\{G^i\}_{i=1}^k$ is obtained as $\mathbf{L}^{i+1} = (\mathbf{P}^i)^T \mathbf{L}^i \mathbf{P}^i$. Matrix $\mathbf{P}^i$ encodes the connection between the nodes of $G^i$ and the nodes of its coarsened version $G^{i+1}$. Briefly, several nodes are aggregated in one, called seed node. The seed nodes of $G^i$ are selected based on their degree and then whether they have a seed neighbor. The rest nodes are aggregated with their closest seeds. Pair of aggregates becomes connected iff any of their constituents were connected.
After coarsening to $G^k$, the uncoarsening process begins. At each level, the current graph $\bar{G}^i$ is edited: some random edges are removed then several new ones are inserted. The same process is performed for the nodes. A newly added node imitates one of the existing nodes, i.e., it copies the structure aggregated within that node. A new edge $(u,v)$ is added by randomly picking a node $u$ and choosing a node $v$, such that the distance between them equals to $d$. Distance $d$ is sampled from the empirical distribution of such distances in the graph $G^i$: for each edge $(u,v)$ the shortest path (except the edge itself) length from $u$ to $v$ is measured. The number of new edges to add is chosen such that to approximately preserve DD. The edited graph version $\tilde{G}^i$ is then uncoarsened to $\bar{G}^{i-1}$.
The editing rates at each level are free parameters, which control the extent of modification and, scaling factor (if the addition rate exceeds the deletion rate). Experiments showed that MUSKETEER was able to reproduce features based on the degree (average degree, assortativity) and distance (average eccentricity, distance, harmonic distance, and betweenness centrality).

In ERGG~\cite{drobyshevskiy2017learning}, the editing occurs at the level of vector representation of its nodes (Figure~\ref{fig:ergg}). An input graph is firstly embedded into vector space, such that a special score function $s(\vec{r}_i, \vec{r}_j)$ is high for edges $i \rightarrow j$ and low for non-edge pairs. The distribution of node vectors $\vec{r}_i \sim \mathcal{R}$ is supposed to encode graph features. New node vectors $\vec{r}'_i$ corresponding to nodes of a new graph are then sampled from $\mathcal{R}$ by resampling known vectors $\vec{r}_i$ and adding small Gaussian noise. Finally, the new nodes are connected according to $s(\vec{r}'_i, \vec{r}'_j)$ values computed for their node vectors. The extent of graph modification could be slightly controlled by the noise magnitude.
Experiments show that ERGG, besides reproducing main graph features, provides variability of graphs, that are generated from one input graph, close to natural variability within a domain~\cite{drobyshevskiy2017reproducing}.

\subsection{Domain-specific class}

Usually, RG models are designed for simple graphs and directed graphs. Some approaches for simple graphs are adapted for directed edges case; one of minor interest is a support of multiple edges, self-loops, etc. They sometimes arise as a byproduct in some models. For example, although SKG takes a simple directed graph as input, its generating process produces multiple edges, which are then removed. Therefore, SKG concept implicitly supports such a property.
Other approaches are not so flexible. For instance, if the edge probability is based on the geometric distance between nodes, one needs additional mechanisms to model self-loops.

In practice, the types of graphs different from the simple one are important. Bipartite graphs which reflect affiliation and authorship networks, attributed graphs where nodes and edges could have labels, and so on.
The domain-specific class is supposed to cover all RG modeling concepts that aimed at producing all types of graphs except simple directed ones. Domain specificity includes: mentioned non-standard edge types, presence of attributes on nodes or edges, special kinds of graphs like bipartite, planar, and so on.

Despite that the defined class is very vast, we consider only two widely used categories: graphs with communities, very popular in the social domain and graphs with weighted edges, widespread in many domains~\cite{barthelemy2005characterization}.

\subsubsection{Community structure}

Complex networks often have groups of more densely connected nodes, called communities. The notion of the community originates from social networks where users unite in groups of common interests, occupation, geography, etc. But community structure also presents in other graph domains. For instance, in protein interaction networks, communities correspond to proteins with similar functionality, in citation networks, nodes group by research topic. Community structure reflects a mesoscale map of network topology and demonstrates its own specific patterns. Last decades, there is considerable interest in community detection methods and in developing accurate models for graphs with community structures.

In this category, we focus on methods for producing explicit community structure in graphs. The concept is to supply each graph node with a label indicating to which communities the node belongs to. Further, we describe models exploiting this concept.
Although we consider this single concept in the category, the approaches for generating graphs with community structure based on this concept could be divided to generative and feature-driven ones\footnote{They are not the subcategories of \cat{community structure} category, because here we consider approaches to creating a \textit{community structure}, while \cat{generative} and \cat{feature-driven} relate to modeling a \textit{graph}.}, according with the described classes.

\paragraph{Generative approaches}

The first step is to define community labels for the nodes. Then, a usual generative pipeline, where edge probability $P_{ij}$ depends on node labels $c_i$ and $c_j$, is applied.

\begin{figure}[h!]
\minipage{0.38\textwidth}
  \includegraphics[scale=0.16]{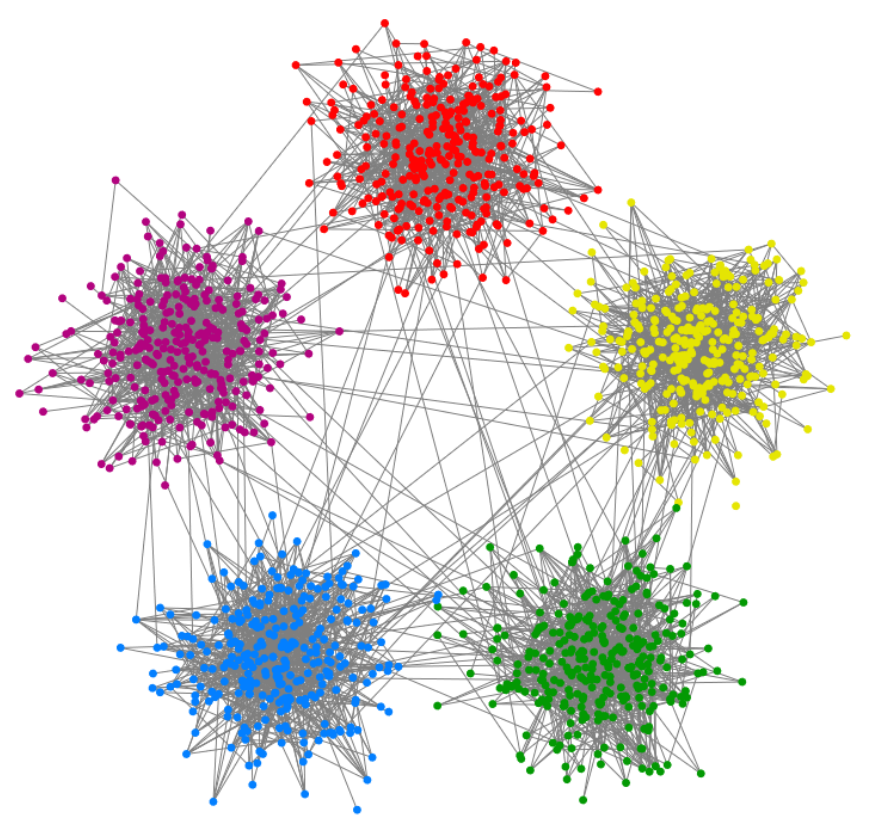}
  \caption{Community structure, modeled as a group of ER graphs with different edge probabilities (Girvan-Newman model, SBM). (Picture from~\cite{abbe2017community})}
  \label{fig:comms}
\endminipage
\hfill
\minipage{0.6\textwidth}
  \includegraphics[scale=0.17]{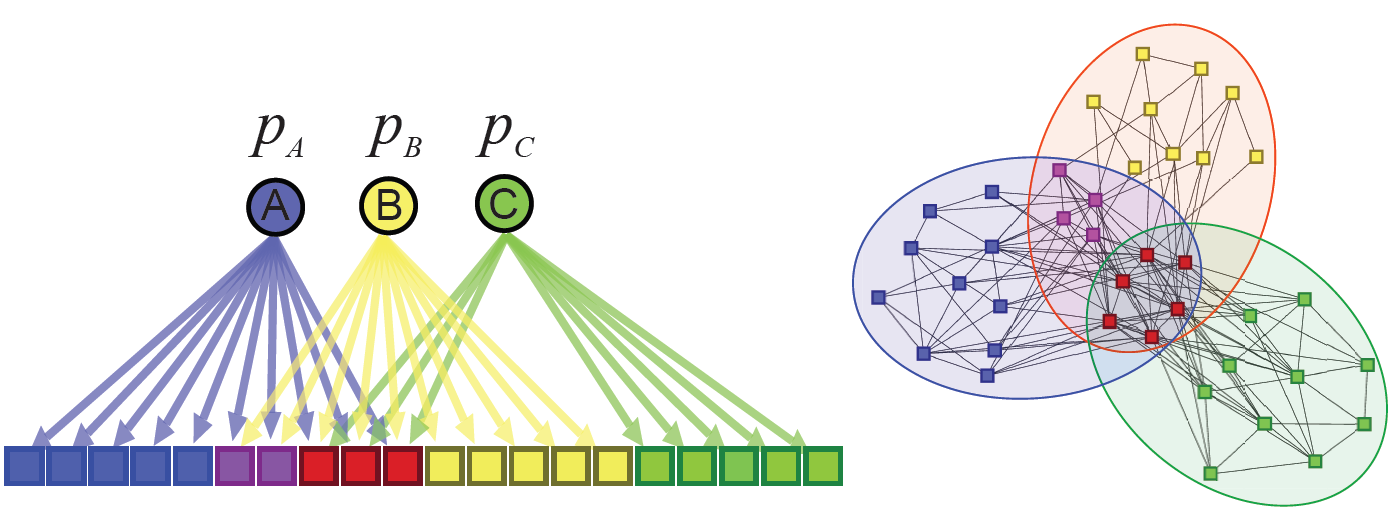}
  \caption{Bipartite affiliation graph determines the association between nodes and communities (AGM, LFR). Picture from \url{http://snap.stanford.edu/agm/} .}
  \label{fig:agm}
\endminipage
\end{figure}

Simplest approaches are based on generating and connecting groups of ER graphs, corresponding to separate communities, with different edge probabilities (Figure~\ref{fig:comms}). Communities could be separate (Girvan-Newman model~\cite{girvan2002community}), intersecting~\cite{sawardecker2009detection}, and could form a hierarchy~\cite{arenas2006synchronization}. In a
BTER model~\cite{seshadhri2012community}, a group of ER blocks is combined with custom node DD. After connecting the nodes within the blocks according to the ER model, "excess" node degrees (equal the wished degree $d_i$ minus real degree within its block, if positive) are used for linking between the blocks using the Expected degree (Chung Lu) model.

A more complex model of assigning nodes to communities is suggested in a Community-affiliation graph model (AGM)~\cite{yang2014structure}. Its first part is a bipartite affiliation graph $B(N, C, M)$, whose edges $M$ indicate to which communities $C$ the nodes $N$ belong to (Figure~\ref{fig:agm}). The second part of AGM defines edge generation model. Probabilities $\{p_c\}$ are defined for each community $c \in C$ and are used to specify the edge probability: $P_{ij} = 1 - \prod_{c \in Z_{ij}} {(1-p_c)}$, where $Z_{ij}$ is a set of common communities for nodes $i$ and $j$.
This model provides important properties of real communities: $P_{ij}$ increases as $Z_{ij}$ increases; the edge density is higher in the intersection of the communities; number of edges $m_c$ in a community $c$ grows super-linearly with its size; community hubs are more likely located in community intersections.
In practice, the affiliation graph $B$ is constructed using a Configuration model, once node membership sequence and community size sequence are specified.

The bipartite affiliation graph is also used in a series of LFR benchmarks~\cite{lancichinetti2009benchmarks}, which provides more flexibility in parameters tuning. A topological mixing parameter $\mu$ is introduced, which controls relative edge density within communities. The internal node degree $d_i^{in}$ is defined as the number of its neighbors sharing at least one community. Its expected value is $d_i^{in} = (1-\mu)d_i$. To achieve this, after forming links within communities, the edge switching procedure is applied.

\paragraph{Feature-driven approaches}

In the feature-driven approaches, node community labels are defined based on a given graph.

SBM~\cite{decelle2011asymptotic} could be employed to fit a real network without ground-truth community structure. It defines a number of groups $Q$, prior group probabilities $p_q$, and a matrix $\mathbf{P}$ of inter-group edge probabilities. The parameters could be estimated, for example, via EM algorithm within the ERGM framework~\cite{decelle2011asymptotic}.
Similar fitting approach is given in the Latent position cluster (LPC) model~\cite{handcock2007model}, which uses the concept of unobserved social space. Communities are represented by a mixture of multivariate normal distributions of points in this space, edge probability depends on euclidean distance: $P_{ij} \sim e^{-\beta dist(i,j)}$. Parameters are then estimated via likelihood maximization or MCMC sampling.

An alternative way is to find communities in an input graph using one of community detection methods and reproduce them in a random graph. In the ReCoN~\cite{staudt2016generating} method, the first step is to detect communities in a given graph. Then the graph with the detected communities is just copied. Finally, the edges are rewired within communities and between the replicas. As a result, the number of communities multiplies by a scaling factor.

In ERGG~\cite{drobyshevskiy2017learning}, an input graph is assumed to have community labels. Due to the mapping of the graph nodes into vectors and sampling new node vectors from the existing ones, the new nodes inherit community labels from the nodes in a proper way. Since labels could attribute to multiple communities, overlapping community structure is supported. As a result, the number of communities remains constant while their sizes change proportionally to a scaling factor.

\subsubsection{Weighted edges}

Edge weights naturally appear in complex networks: they could express the strength of ties in a social network, flux amount in a metabolic reaction, gene co-expression measure, etc. Multiple edges in the graph could also be interpreted as integer weights. Many metrics and concepts generalize to weighted graphs, including shortest path length, clustering, modularity measures. Considering of weighted graphs brings new aspects to the existing network tasks such as community detection~\cite{newman2004analysis}.

One way to get a weighted edge is to treat multi-edges as weighted ones. In RTG~\cite{akoglu2009rtg} model, based on a random character sequence, each next repetition of the  same pair of words increments the corresponding edge weight. It leads to the power law of node strength dependence on its degree: $s_i \sim d_i^{\beta}$. The RTG algorithm also provides the total weight power law, $W(t) \sim m(t)$, and self-similar weight addition.

In the LFR~\cite{lancichinetti2009benchmarks} benchmark, the weights are assigned to the edges, such that for each node $i$, the expected node strength is $s_i \sim d_i^{\beta}$. Node internal strength $s_i^{in}$ (the strength computed for community neighbors only) is controlled by a mixing parameter $\mu$: $s_i^{in} = (1-\mu)s_i$. These conditions are achieved by a greedy algorithm which iteratively modifies the edge weights $w_{ij}$ in order to minimize the quadratic variance of all $s_i$, $s_i^{in}$, and $s_i - s_i^{in}$ summed up over all the nodes.

\section{Discussion}
\label{sec:discussion}

\begin{table}[h!]
\caption{Popular random graph models (rows) combining concepts (columns) from several classes --- marked as '\checkmark' in corresponding cells. Colors correspond to classes: blue for Generative, green for Feature-driven, red for Domain-specific.}
\begin{center}
\begin{tabular}{|l|b|b|b|b|b|b|b|b|g|g|g|g|g|r|r|}
\hline
& 
\rotatebox[origin=l]{90}{simple (ER)} & 
\rotatebox[origin=l]{90}{PA} & 
\rotatebox[origin=l]{90}{copying} & 
\rotatebox[origin=l]{90}{other local rules} & 
\rotatebox[origin=l]{90}{recursive} & 
\rotatebox[origin=l]{90}{geometry} & 
\rotatebox[origin=l]{90}{node labeling} & 
\rotatebox[origin=l]{90}{topology from optimization} & 
\rotatebox[origin=l]{90}{analytical} & 
\rotatebox[origin=l]{90}{parameters estimation} & 
\rotatebox[origin=l]{90}{ERGM} & 
\rotatebox[origin=l]{90}{edge switching} & 
\rotatebox[origin=l]{90}{other editing} & 
\rotatebox[origin=l]{90}{community structure} & 
\rotatebox[origin=l]{90}{weighted edges} \\ \hline
R-MAT~\cite{chakrabarti2004r} &  &  &  &  & \checkmark &  & \checkmark &  &  &  &  &  &  &  &  \\ \hline
SKG~\cite{leskovec2010kronecker}
&  &  &  &  & \checkmark &  & \checkmark &  &  & \checkmark &  &  &  &  &  \\ \hline
MFNG~\cite{palla2010multifractal} &  &  &  &  & \checkmark &  & \checkmark &  &  & \checkmark &  &  &  &  &  \\ \hline
RTG~\cite{akoglu2009rtg} &  &  &  &  & \checkmark &  & \checkmark &  &  &  &  &  &  &  & \checkmark \\ \hline
Forest Fire~\cite{leskovec2007graph} &  &  & \checkmark & \checkmark & \checkmark &  &  &  &  &  &  &  &  &  &  \\ \hline
GScaler~\cite{zhang2016gscaler} &  &  & \checkmark & \checkmark &  &  &  &  &  &  &  &  &  &  &  \\ \hline
MUSKETEER~\cite{gutfraind2015multiscale} &  &  & \checkmark & \checkmark & \checkmark &  &  &  &  &  &  &  & \checkmark &  &  \\ \hline
LPC~\cite{handcock2007model} &  &  &  &  &  & \checkmark & \checkmark &  &  & \checkmark &  &  &  & \checkmark &  \\ \hline
Girvan-Newman~\cite{girvan2002community} & \checkmark &  &  &  &  &  & \checkmark &  &  &  &  &  &  & \checkmark &  \\ \hline
BTER~\cite{seshadhri2012community} & \checkmark &  &  &  &  &  & \checkmark &  & \checkmark &  &  &  &  & \checkmark &  \\ \hline
AGM~\cite{yang2014structure} &  &  &  &  &  &  & \checkmark &  & \checkmark &  &  &  &  & \checkmark &  \\ \hline
LFR~\cite{lancichinetti2009benchmarks} &  &  &  &  &  &  &  &  & \checkmark &  &  & \checkmark &  & \checkmark & \checkmark \\ \hline
ReCoN~\cite{staudt2016generating} &  &  & \checkmark &  &  &  &  &  &  &  &  & \checkmark &  & \checkmark &  \\ \hline
SBM~\cite{decelle2011asymptotic} & \checkmark &  &  &  &  &  &  &  &  & \checkmark & \checkmark &  &  & \checkmark &  \\ \hline
ERGG~\cite{drobyshevskiy2017learning} &  &  & \checkmark &  &  & \checkmark &  &  &  & \checkmark &  &  & \checkmark & \checkmark & \checkmark \\ \hline
GIRG~\cite{bringmann2016geometric} &  &  &  &  &  & \checkmark &  &  & \checkmark &  &  &  &  &  &  \\ \hline
Dot-product~\cite{nickel2008random} &  &  &  &  &  & \checkmark & \checkmark &  &  &  &  &  &  &  &  \\ \hline
BRITE~\cite{medina2000origin} &  & \checkmark &  &  &  & \checkmark &  &  &  &  &  &  &  &  &  \\ \hline
\end{tabular}
\end{center}
\label{tab:models_ideas}
\end{table}

In this section, we discuss the taxonomy presented in the previous section and outline how it works at various RG applications.

\subsection{Taxonomy discussion}

\paragraph{Relation of concepts and models}
If we tried to build a taxonomy of RG models, it would be a huge branching tree, where similar concepts would repeat many times. Moreover, it is hard to classify models themselves, since they often mix different approaches. This is why some models appear in several categories of the taxonomy.

Our taxonomy presents and classifies the main concepts used in the RG models. We consider how the models combine these concepts. For that, we compare the models, based on two or more concepts, in a Table~\ref{tab:models_ideas}.
One can see that models can exploit up to six concepts in various combinations. ERGG~\cite{drobyshevskiy2017learning} is an algorithm which uses parameter estimation technique to learn the geometrical representation, then the copying and editing mechanisms to scale and randomize graph, and produces community labels and edge weights (although just using copying again).
Forest Fire~\cite{leskovec2007graph} model employs three generative mechanisms together: copying in- and out-links of a chosen node, the local rules while determining the unvisited neighbors to decide where to proceed the burning process, and the recursive principle when repeating the same procedure at each node.

\begin{figure}[h!]
\centering
\includegraphics[scale=0.5]{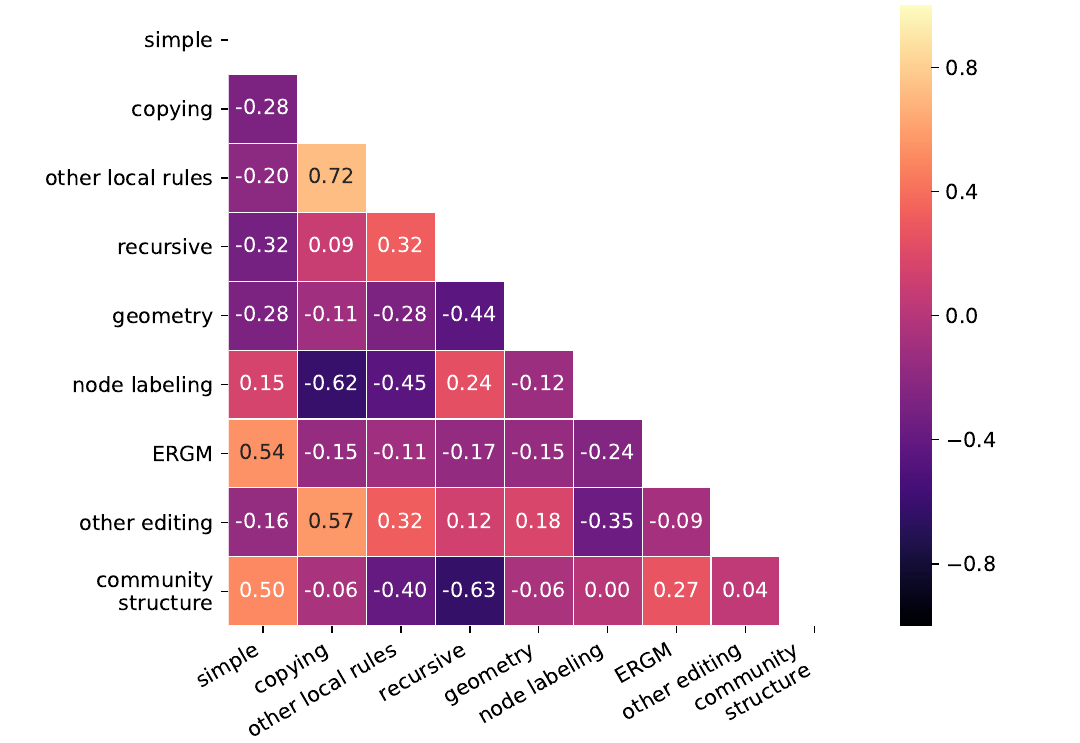}
\caption{Correlations of several concepts co-occurrence in RG models. Computed based on table~\ref{tab:models_ideas}. Categories with all correlations close to 0 are excluded.}
\label{fig:corr}
\end{figure}

Although the table is small, we computed correlations between its columns (marked cells were treated as ones, empty cells were treated as zeros). Figure~\ref{fig:corr} represents the correlation matrix for categories that show the highest or lowest correlations.
High correlation means that concepts are well compatible. Low correlation means the opposite. One can see that the \cat{copying} concept is mixed well with \cat{other local rules} (0.72) and \cat{other editing} (0.57), which could correspond to the evolutionary principle of copying with mutations. 
\cat{Simple} approach is popular within \cat{ERGM} framework (0.54 and for creating community structure (0.50).

Being at the opposite side of the list, low correlations may indicate that concepts are incompatible or were not used together for some reason. \cat{Community structure} combines poorly with \cat{recursive} approaches (-0.63), which could seem strange, because it is known that recursive graph structure is related to its hierarchical community structure. However, the explanation could be that current recursive generative mechanisms (recall SKG, RTG, Forest Fire) fail to produce an explicit community structure with the desired features. Perhaps, the bottleneck is parameters fitting, because both graph features and community features must be fitted. SKG and MFNG already have intricate fitting procedures, satisfying all the conditions is too complex.
The next elements with the lowest correlations are \cat{node labeling} with \cat{copying} (-0.62) and \cat{other local rules} (-0.45), --- are harder to explain. Perhaps, such combinations exist but are under-represented in our collection, or wait to be discovered.

\paragraph{Choice of categories and generality of the taxonomy}
The number of classes in the taxonomy is somewhat arbitrary. While the Generative and Feature-driven classes are intended to cover all models of simple and directed graphs, the lowest level categories still unite many approaches and could be specified further. For instance, recursive approaches can be deterministic or stochastic, geometric approaches may contain a subcategory \cat{hyperbolic geometry} due to a large number of works devoted to hyperbolic random graphs. On the other hand, the RG models often use a lot of various heuristics that are hard to classify and potentially could form their subcategories. For example, refer to \cat{other local rules} section, where rules of neighbor choosing or triadic formation could be differentiated. Therefore, we chose three levels of abstraction as a kind of a compromise for concepts granularity.

Several categories, namely \cat{other local rules}, \cat{other editing}, and \cat{node labeling} serve as containers for approaches not present in other categories of that class. For example, \cat{node labeling} corresponds to those methods based on node attributes that are not geometric --- there are not so many concepts, according to our knowledge.

Speaking about the taxonomy generality, we designed the three classes to cover the main directions of RG modeling. Generative and Feature-driven classes are described in details and well-structured. We assume that they reflect the state-of-affairs in modeling simple and directed graphs. In the domain-specific class we include only 2 most popular cases and decided to leave the rest out of scope. Since there exist many other specific types of graphs, the third class structure is far from complete.

New concepts, as well as those we missed, are supposed to fall into an existing (sub)category or form a new one in one of the classes. Regarding the emergence of new concepts, it seems that generative class of approaches exhausts itself. Main network formation mechanisms are already invented and described in the literature. The further progress is expected from the feature-driven approaches. The main challenges of RG modeling concern better fitting a model to a given graph, creating of fast and simple procedures of graph sampling.
Perspective future directions could be: graph editing based on graph representation learning; methods for generation of very large graphs with billions of edges.

Finally, we suppose that clarifying of the concepts that proved their workability in the RG modeling will promote the development of new models. However, a new model is not merely a mix of several concept stubs, it is usually aimed at answering a practical challenge. Now we discuss which concepts are successful at which tasks.

\subsection{Applications of random graph modeling}
\label{ssec:applications}

We identify six directions, where RG models have their applications: networks understanding, analysis, extrapolating, benchmarks, null models, and randomization. Further, we show how the concepts, described in the previous section, are applied to solve problems occurring in these areas.
The results are generalized in table~\ref{tab:applications_models}.

\subsubsection{Understanding}

Discovering new topological patterns in real networks posed a need to explain their emergence. If a hypothesized generative mechanism produces graphs with the same patterns, it could underlie the real processes of the network formation.
Therefore, all concepts from Generative class are potential explanations of network formation. The preferential attachment rule, being coupled with nodes addition, generates the scale-free topology, has intuitive interpretations. A new person joining a social network more likely makes a connection to a hub. A new web page is more likely to link to a page with many links. The same for scientific papers citing. Another well demonstrative example is copying principle. Copying edges of a node corresponds to inheriting citation links in citation networks and WWW, genes duplication in biological networks, and so on. Node attributes based linking is consistent with the homophilic attraction of similarities. A recursive procedure is connected to self-similarity, and its workability could indicate that the same simple laws govern networks formation at different scales. Works on obtaining topologies resulting from optimization tasks, evidence that network structures emerge in a way to be optimal in some sense.

Feature-driven statistical models such as ERGMs also contribute to the understanding of complex networks by the following reasons. ERGM framework is used to test, how various graph statistics could explain the observed structure. The model that fits best (in some sense) to the real network indicates what features are most important to explain the network architecture~\cite{simpson2011exponential}.
The stochastic model can capture not only regularities in a graph, but also variabilities of its properties. Being fitted to a real graph, the model gives a picture of the distribution of possible observable outcomes~\cite{robins2007introduction}.
Several features could have more than one explanation. For instance, the high CC could be caused by homophily or could emerge from self-organizing structural effects. A model combining both effects helps to estimate the contributions of both alternatives quantitatively~\cite{robins2007introduction}.

\subsubsection{Analysis}

Simple models, such as ER with edge probability depending on graph size $p(n)$, were deeply studied on their evolution behaviour, i.e., when $n$ tends to infinity. Various kinds of phase transitions were discovered, e.g., the emergence of a giant connected component and triangles~\cite{raigor2012models}. It motivated to study robustness of networks like the Internet, communication nets, etc., in terms of resilience to attacks like random or intended removing of nodes or edges~\cite{callaway2000network}.

In order to analyze processes taking place on networks: the spread of information or epidemics in social networks; flows in transportation networks and the Internet; economic transactions, and so on, one needs to perform simulation studies. Since the topology of interactions is crucial for processes dynamics, therefore a need for realistic graph models~\cite{castellano2009statistical}.
According to review of publications in JASSS (1998 -- 2015) Fr{\'e}d{\'e}ric Amblard et al.~\cite{amblard2015models}, the majority of works actually use very simple models: regular lattices, random graphs, small-world networks or scale-free network.
Authors suggest three perspectives for social network models:
\begin{itemize}
\item Abstract models, like Forest Fire~\cite{leskovec2007graph}, reproducing a lot of the known properties. These correspond to our Generative class of approaches. Benchmarks like LFR~\cite{lancichinetti2009benchmarks} are promising to employ community structure for populations modeling.
\item Models able to fit a given network sample, e.g., ERGMs and SKG~\cite{leskovec2010kronecker}. In our case, this concept is described in the \cat{fitness optimization} category.
\item Rule-based approach, where a population is generated using Bayesian rules and then a graph is constructed by specifying nodes matching rules, for example in work of S.\,Thiriot and J.-D.\,Kant~\cite{thiriot2008generate}. This approach is reflected in categories \cat{latent attributes} and \cat{local rules}.
\end{itemize}

\begin{table}[t]
\caption{How concepts (columns) from the taxonomy work in six random graph application directions (rows), described in section~\ref{ssec:applications}. If the application area involves RG models employing the concept, the corresponding cell is marked as '\checkmark'. Colors correspond to taxonomy classes: blue for Generative, green for Feature-driven, red for Domain-specific.}
\begin{center}
\begin{tabular}{|l|b|b|b|b|b|b|b|b|g|g|g|g|g|r|r|}
\hline
& 
\rotatebox[origin=l]{90}{simple (ER)} & 
\rotatebox[origin=l]{90}{PA} & 
\rotatebox[origin=l]{90}{copying} & 
\rotatebox[origin=l]{90}{other local rules} & 
\rotatebox[origin=l]{90}{recursive} & 
\rotatebox[origin=l]{90}{geometry} & 
\rotatebox[origin=l]{90}{node labeling} & 
\rotatebox[origin=l]{90}{topology from optimization} & 
\rotatebox[origin=l]{90}{analytical} & 
\rotatebox[origin=l]{90}{parameters estimation} & 
\rotatebox[origin=l]{90}{ERGM} & 
\rotatebox[origin=l]{90}{edge switching} & 
\rotatebox[origin=l]{90}{other editing} & 
\rotatebox[origin=l]{90}{community structure} & 
\rotatebox[origin=l]{90}{weighted edges} \\ \hline
understanding  & \checkmark & \checkmark & \checkmark & \checkmark & \checkmark & \checkmark & \checkmark & \checkmark &   &   & \checkmark &   &   &   &  \\ \hline
analysis  & \checkmark & \checkmark & \checkmark & \checkmark & \checkmark & \checkmark & \checkmark &   & \checkmark & \checkmark & \checkmark & \checkmark &   & \checkmark & \checkmark \\ \hline
extrapolating  &   &   & \checkmark & \checkmark & \checkmark & \checkmark & \checkmark &   &   & \checkmark &   & \checkmark & \checkmark & \checkmark & \checkmark \\ \hline
benchmarks  &   & \checkmark & \checkmark &   &   & \checkmark &   &   &   & \checkmark & \checkmark & \checkmark &   & \checkmark & \checkmark \\ \hline
null models  & \checkmark &   &   &   &   &   &   &   & \checkmark & \checkmark & \checkmark & \checkmark &   & \checkmark &  \\ \hline
randomization  &   &   &   &   &   &   &   &   & \checkmark &   &   & \checkmark & \checkmark &   &  \\ \hline
\end{tabular}
\end{center}
\label{tab:applications_models}
\end{table}

\subsubsection{Extrapolating}

The Internet, social networks, and other graphs grow over time, which raises questions, like `will the Internet protocol work over five years?', or `how will Facebook look like in future?'
To answer these questions, one needs algorithms for scaling an existing graph to a larger size. Moreover, graphs of interest are themselves already large, billions of nodes of order, which impose severe constraints on the algorithms complexity and memory usage.
The problem is generally addressed in two ways: a generative process capturing the needed features and producing graphs of custom size or a method for scaling up a given graph.

From the Generative class, \cat{recursive}, \cat{latent attributes} and \cat{copying} concepts have the most success. Recursive matrix multiplication (SKG~\cite{leskovec2010kronecker}, MFNG~\cite{palla2010multifractal}, TrillionG~\cite{park2017trilliong}) is elegant but needs a fast algorithm for graph construction, since $O(n^2)$ generation time may be impractical in case of very large graphs. Moreover, recursion in these algorithms often implies too discrete graph size $n = n_0^k$ which needs some workaround.
Copying principle is employed at different levels: from replication of nodes with edge stubs (GScaler~\cite{zhang2016gscaler}) to copying of whole communities and graph itself (ReCoN~\cite{staudt2016generating}), and copying vector representations of nodes (ERGG~\cite{drobyshevskiy2017learning}).
Imitating of features of the graph, normally requires either model parameters fitting (SKG, MFNG, ERGG), or special heuristics to tune construction process towards the desired properties (GScaler).
However, the other approaches from \cat{latent attributes} category like hyperbolic or dot-product graphs are promising candidates. Perhaps, they just lack their fitting procedures.

A non-generative way of scaling up a graph assumes its modification, increasing the graph size. For instance, a series of coarsening and then a longer series of uncoarsening procedure leads to a larger version of the graph (MUSKETEER~\cite{gutfraind2015multiscale}). In a more naive case, a graph with communities is multiplied and its edges are rewired (ReCoN). Note that both examples are based on copying.

\subsubsection{Benchmarks}

There is a wide range of network mining tools and algorithms being actively developed in various network domains. Examples vary from metric calculators (like diameter, modularity) to topology inference (link prediction, etc.)~\cite{newman2001random, barrat2008, alessandro2007large}. Reliable testing of these methods is complicated or even impossible due to the lack of proper test data. For instance, significance testing requires a representative set of graphs providing realistic variability of their features, while for scalability testing, one needs a series of similar graphs of different sizes~\cite{drobyshevskiy2017learning}.
The answer could be benchmark random graphs tailored for specific cases.

The representativeness of the sample is of main importance for significance testing.
A perfect solution could be ERGMs, since they define graph probability space with theoretically any constraints on its statistics. Unfortunately, in practice, current approaches suffer from serious computational issues.
Controllable edge switching within MCMC sampling framework can be used to achieve graphs with desired feature ranges and feature distribution constraints~\cite{ying2009graph}, while the degree sequence remains constant.
Less strict in features, but providing controllable variability of generated graphs, is a recursive SKG model with dependent edge sampling~\cite{morenomodeling}. The same for the copying based ERGG~\cite{drobyshevskiy2017learning}, where the magnitude of noise added to the learned node vectors, encodes variability of result graphs.

Scalability testing implies a generator of graphs of controlled size. If similarity to a given graph is required, the problem reduces to the extrapolating task, possibly with a scale factor less than 1. If the particular topology is not important, any algorithm from the Generative class, capable of producing features of interest, is of use.
Computational and memory issues arise here when the size of graphs become very large. Quadratic time algorithms are impractical with over a million nodes. Acceptable complexity is $O(n \log n)$ or linear to the number of edges $O(m)$, which means that some techniques, such as pairwise nodes matching, are eliminated.
Solutions for large graphs could be to adapt existing algorithms (ROLL~\cite{hadian2016roll}: speed up of PA), approximate them, or develop their distributed versions~\cite{meyer2017large}. Distributed algorithms allow to achieve graph with billions of nodes in CKB~\cite{chykhradze2014distributed} and trillions of edges in Darwini~\cite{edunov2016darwini}.

Of special interest are random graphs with communities, especially for social networks emulating. The analytical approach is employed to implement a rich set of parameters describing DD and community features. The Configuration model for realizing DD and edge switching for randomization is used in the LFR benchmark~\cite{lancichinetti2009benchmarks}. Distributed algorithms are also actual here~\cite{chykhradze2014distributed, edunov2016darwini}.

\subsubsection{Null models}

Null models serve for accurate hypothesis testing, providing a dataset with well controllable features. In the context of complex networks, one can learn what patterns are specific to the given network, compared to some common distribution. The null model allow quantifying how much the network is different from a random one (null hypothesis). For example, to evaluate the statistical significance of the edge reciprocity or the observed number of common neighbors of two nodes~\cite{zweig2016network}.

The most popular random graph null model is the Configuration model which specifies a uniform distribution over all graphs with the same degree sequence. It is widely employed to explore network patterns in sociology, ecology, systems biology and other domains, refer to B.\,K.\,Fosdick~\cite{fosdick2018configuring} for a review. For example, detecting network motifs as over-represented subgraphs~\cite{milo2004superfamilies}, or serving as a null model against which the modularity is measured~\cite{newman2004finding}.
Uniform sampling of graphs with fixed degree sequence is efficiently performed via MCMC sampling techniques based on the~\cite{fosdick2018configuring}.

An alternative null model could be Chung-Lu model which sets the expected node degrees $d_i$ instead of fixed ones, with the edge probability $P_{ij} \sim d_id_j$. Its use is related to a community detection method based on SBMs~\cite{karrer2011stochastic}.

\subsubsection{Randomization}

Sharing graph datasets can be problematic if a graph contains private information. There is a need to randomize the graph in a way to avoid the leak of confidential information and at the time to preserve its important topological features.
It was shown that the simple relabeling of nodes identifiers does not guarantee the safety~\cite{backstrom2007wherefore}. Possible attacks include active and passive attacks. An active attack could be the introducing of 'sybil' nodes with a specific distinguishable configuration into a prior graph in order to discover it after randomization. A passive attack can occur when a user could de-anonymize himself using knowledge of near network topology or some auxiliary information like aggregated social networks~\cite{narayanan2009anonymizing}.

Generally, RG models can help the anonymization in two ways: generate a graph, thoroughly repeating the patterns of the original one, or randomize the origin via graph editing methods.
Although generative algorithms face the fitting problem, they are potentially more reliable since the mapping between original nodes and generated nodes is absent. Anonymization corresponds to extrapolation with scaling factor of 1. Therefore, all solutions could potentially be employed.

However, a more widespread technique is randomization via graph editing. Edge switching is fast and easy to implement, although it violates all features except degree sequence. Nevertheless, edge switching under constraints (e.g., preserving chosen spectral features), combined with random edge addition and deletion, is claimed to protect edge privacy~\cite{ying2008randomizing}.
The $dK$-random graph model is also used to capture graph patterns. Measured $dK$-distributions are then perturbed  to enforce differential privacy~\cite{dwork2008differential} guarantees, and a new graph is generated according to the new $dK$-distributions~\cite{wang2013preserving}.

Following the review of Shouling Ji et al.~\cite{ji2015secgraph}, other randomization approaches include $k$-anonymity (where each node in the graph to be published has $k-1$ symmetric nodes), cluster-based anonymization (graph structure is preserved at the level of clusters ignoring their inner configuration), Random Walk based anonymization (edge $(u,v)$ is replaced with $(u,w)$ where $w$ is destination of random walk from $u$), and others.

\section{Summary}
\label{sec:conclusion}

In the survey, we presented a novel view on random graph modeling approaches. We detected main concepts used in RG models and organized them into a hierarchical taxonomy, consisting of three classes: Generative, Feature-driven, and Domain-specific.
We hope that the taxonomy of concepts will help researchers to orient in an enormous amount of existing RG models and develop their models based on the experience of previous work.
Although these classes cover existing approaches, we considered only two categories in the Domain-specific class. Due to a wide variety of graph types, this class could be significantly extended.
An interesting and promising direction for future work is the deep neural networks for RG modeling.

\bibliographystyle{acm}
\bibliography{main}

\end{document}